\documentclass[structabstract]{aa} % for a referee version
\usepackage{graphicx}
\usepackage{txfonts}

\def\mjup{M$_{\rm Jup}$}

\def\bp{$\beta$~Pic\ }

\def\L'{m$_{L'}$\ }

\def\pm{$^{+}_{-}$}
\def\deg{$^{\circ}$}
\def\mjup{M$_{\rm Jup}$\ }

\begin{document}

\title{An insight in the surroundings of HR4796\thanks{Based on observations collected at the
European Southern Observatory, Chile, ESO; run 085.C-0277A.}}
%\subtitle{Constrains on their relative position }

\author{
A.-M.~Lagrange \inst{1}
\and
J. Milli \inst{1}
 \and
A. Boccaletti \inst{2}
\and
S. Lacour \inst{2}
\and
P. Thebault \inst{2}
\and
G. Chauvin \inst{1,3}
\and
D. Mouillet \inst{1}
\and
J.C. Augereau \inst{1}
\and
M. Bonnefoy \inst{3}
\and
D. Ehrenreich \inst{1}
\and
Q. Kral \inst{2}
}

\offprints{A.-M. Lagrange}

\institute{
  Institut de Plan\'etologie et d'Astrophysique de Grenoble,
  Universit\'e Joseph Fourier, CNRS, BP 53, 38041 Grenoble, France
  \email{anne-marie.lagrange@obs.ujf-grenoble.fr}
\and
LESIA-Observatoire de Paris, CNRS, UPMC Univ. Paris 06, Univ. Paris-Diderot, 92195, Meudon, France 
\and
Max Planck Institut fur Astronomie K\"onigstuhl 17, D-69117 Heidelberg, Germany
}

\date{Received date / Accepted date}

\abstract
{  HR4796 is a young, early A-type star harbouring a well structured debris disk, shaped as a ring with sharp inner edges. The inner edge might be shaped by a yet unseen planet inside the ring; the outer one is not well understood. The star forms together with the M-type star HR4796B, a binary system, with a projected separation of $\simeq$ 560 AU.}
{Our aim is to explore the surroundings of HR4796A and B, both in terms of extended or point-like structures.}
{Adaptive optics images at L'-band were obtained with NaCo in Angular Differential Mode and with Sparse Aperture Masking (SAM). We analyse the data as well as the artefacts that can be produced by ADI reduction on an extended structure with a shape similar to that of HR4796A dust  ring. We determine constraints on the presence of companions using SAM and ADI on HR4796A, and ADI on HR4796B. We also performed dynamical simulations of a disk of planetesimals and dust produced by collisions, perturbed by a planet located close to the disk outer edge.}
{The disk ring around HR4796A is well resolved. We highlight the potential effects of ADI reduction of the observed disk shape and surface brightness distribution, and side-to-side  asymmetries. We produce 2D maps of planet detection limits. No planet is detected around the star, with masses as low as 3.5 \mjup at 0.5" (58 AU) and less than 3 \mjup in the 0.8-1" range along the semi-major axis. We exclude massive brown dwarfs at separations as close as 60 mas (4.5 AU) from the star thanks to SAM data. The detection limits obtained allow us to exclude a possible close companion to HR4796A as the origin of the offset of the ring center with respect to the star; they also allow to put interesting constraints on the (mass, separation) of any planet possibly responsible for the inner disk steep edge.  Using detailed dynamical simulations, we show that a giant planet orbiting outside the ring could sharpen the disk outer edge and reproduce the STIS images published by Schneider et al. (2009). Finally, no planets are detected around HR4796B with limits well below 1 \mjup at 0.5" (35 AU).}
{}


\maketitle

\section{Introduction}
Understanding planetary systems formation and evolution has
become one of the biggest challenges of astronomy, since the imaging of a debris disk around $\beta$ Pictoris in
the 80's (\cite{smith84}) and the discovery of the first exoplanet around the solar-like
star 51 Pegasi during the 90's (\cite{mayor95}). Today, about 25 debris disks have been
imaged at optical, infrared, or submillimetric wavelengths (http://astro.berkeley.edu/kalas/disksite). 
Debris disks trace stages of system evolution where solid bodies with sizes significantly larger than the primordial dust size (larger than meters or km sized) are present to account for through collisions, the presence of short lived dust. They are thought to be privileged places to search for planets. This is particularly true  for those showing peculiar structures (e.g. rings with sharp edges) or asymmetries, spirals even though other physical effects not involving planets could also lead to the formation of similar structures. \cite{takeuchi01} for instance showed that relatively small amounts (typ. 1-a few Earth masses) of gas can shape the dust disk through gas-dust interactions into rings (see below). It is remarkable however that all the stars around which relatively close (separations less than 120 AU) planets have been imaged are surrounded by debris disks: a $\leq$ 3-\mjup planetary companion was detected in the outskirts of Fomalhaut's debris disk (119 AU from the star; \cite{kalas08}), four
planetary companions of 7-10 \mjup were imaged at 15, 24, 38 and 68 AU (projected separations) from HR 8799
(\cite{marois08}, \cite{marois10}). Using VLT/NaCo L'-band saturated images, we detected a 9\pm 3 \mjup planet in the disk of $\beta$ Pictoris ($\simeq$ 12 Myr) with an orbital radius of 8-12 AU from the star (\cite{lagrange10}; 
\cite{chauvin12}). More recent studies at Ks show that $\beta$ Pic b is
located in the inclined part of the disk (\cite{lagrange11}), conforting the link between disk morphology and the
presence of a planet (\cite{mouillet97}; \cite{jca01}). $\beta$ Pic b also confirms that giant planets form in a timescale of 10 Myr or less. Interestingly, $\beta$ Pic b and maybe also HR8799e could have formed in situ via core accretion, in contrast with the other, more remote, young companions detected with high-contrast imaging. If formed in situ, the latter probably formed through gravitational instabilities within a disk, or through the fragmentation and collapse of a molecular cloud. 

There are many exciting questions regarding disks and planets: could different planet formation processes be at work within a given disk? Disks and planets are known to exist in binary systems; (how) do massive companions impact on the dynamical evolution of inner planets and disks?  In a recent study, \cite{rodriguez11} showed
%of more than 115 debris disks around main sequence stars, \cite{rodriguez11} showed that 25\pm 4 $\%$ of debris disk systems are binary or triple star systems; they also 
 that for a binary system to have a disk, it must either be a very wide binary system with disk particles orbiting a single star or a small separation binary with a circumbinary disk. Such results can help the search for planets if one relates debris disks with planet formation. However, another question, already mentioned, is to which extent and how can debris disks indicate the presence of already formed planets? 

A particularly interesting system in the present context is HR4796A, consisting of an early-type (A0), young, close-by star (see Table~\ref{star})  
surrounded by dust, identified in the early 90's (\cite{jura91}) and resolved by \cite{koerner98} and \cite{jaya98} at mid-IR from the ground, and at near-IR with NICMOS on the HST (\cite{schneider99}) as  well as from the ground, coupling coronagraphy with adaptive optics (\cite{jca99}). The resolved dust shapes as a narrow ring, with steep inner and outer edges. The steepness of the  inner edge of the dust ring has been tentatively attributed to an unseen planet (\cite{wyatt99}); however, none has been detected so far. The disk images + SED modeling required at least two populations of grains, one, narrow (a few tens AU) cold ring, located at $\simeq$ 70 AU, and a second one, hotter and much closer to the star (\cite{jca99}), although the existence of an exozodiacal dust component is debated (\cite{li03}). \cite{wahhaj05}, argued that in addition to the dust responsible for the ring-like structure observed at optical/near IR wavelengths, a wider, low-density component should be present at similar separations to account for the thermal IR images. Recently, higher quality (SN/angular resolution) data were obtained with STIS (\cite{schneider09}) and from the ground (\cite{thalmann11}), the latter using  performant AO system on a 8 m class telescope, as well as Angular Differential Imaging (ADI, see below). With the revised distance of HR4796 with respect to earlier results, the ring radius is at about 79 AU, and has a width of 13 AU (STIS 0.2-1 $\mu$m data). Furthermore, these authors show a 1.2-1.6 AU physical shift of the projected center of the disk wrt the star position along the major axis and \cite{thalmann11} moreover measures a 0.5 AU shift along the disk minor axis. 
Finally, and very interestingly in the context of planetary system formation, HR4796 has also a close-by (7.7 arcsec, i.e. about 560 AU projected separation), M-type companion (\cite{jura93}), plus a tertiary one, located  much further away (13500 AU in projection; \cite{kastner08}). The closest companion may have played a role in the outer truncature of the disk, even though, according to \cite{thebault10}, it alone cannot account for the sharp outer edge of HR4796 as observed by STIS.

 \begin{table}[t!]
    \caption{HR 4796 stellar properties. (1)\cite{vanleeuwen07}, (2) \cite{stauffer95}. }
   \label{star}
 \begin{center}
      \begin{tabular}{l l}
	\hline
        Spectral Type                       & A0V \\
        $V$                                & 5.78 \\
        $B-V$                               & 0.009\\
        $\pi$                         &  3.74 \pm 0.33 mas (1)\\
	Age                             & 8\pm 2 Myr (2) \\
        \hline
      \end{tabular}
    \end{center}
\label{star}
  \end{table}

ADI (\cite{marois06}) is a technics that has proved to be very efficient in reaching very high contrast from the ground on point-like objects. It has also been used to image disks around HD61005 (the Moth disk, \cite{buenzli10}), HD32297 (\cite{bocca11}), and \bp (\cite{lagrange11}), but it should be used with care when observing extended structures, as the morphology of these structures may be strongly impacted by this method  (Milli et al, 2012, in prep). 

 We present here new high contrast images of HR4796 obtained with NaCo on the VLT at L' band, both in ADI and Sparse Aperture Mode (SAM, \cite{tuthill10}) aiming at exploring the disk around HR4796A, and at searching for possible companions around HR4796A as well as around HR4796B. SAM and ADI are complementary as the first give access to regions in the 40-400 mas range from the star, and ADI further than typically 300 mas. Pedagogical examples of SAM performances and results are reported in \cite{lacour11}.

\section{Data log and Reduction}
\subsection{Log of observations}
VLT/NaCo (\cite{lenzen03}; \cite{rousset03}) L'-band data were obtained on April, 6 and 7th, 2010, in ADI mode and with SAM. We used the L27 camera which provides a pixel scale of $\simeq$ 27 mas/pixel. 
The log of observations is given in Table~\ref{logobs}.

 \begin{table*}[t!]
  \caption{Log of observations. 
 ``Par. range'' stands for the parallactic angles at the start and end of 
observations; 
``EC mean'' for the average of the coherent energy and ``t0 mean'' for the average of the coherence time during the observations. }
\label{stats}
\begin{center}
\begin{tabular}{ l l l l l l l l l}
\hline 
 Date & UT-start/end&DIT & NDIT & N exp. & Par. range & Air Mass range& EC mean & t0
mean  \\ 
	&(s) &     &     &	& \deg   & &$\%$& (ms)\cr
\hline 
Apr. 07, 2010 & 03:08/03:13     &0.2 (ND Long) &150 &8    &-48.3 to-47.2 &1.07  &44.9&3.0\\ 
Apr. 07, 2010 & 03:14/04:56     &0.2           &150 &160  &-45.3 to 31.9  &1.07 to 1.05  &41.4&2.6 \\
 Apr. 07, 2010 & 04:57/05:06     &0.2 (ND Long) &150 &8    &32.6 to 38.3 &1.05 &44.0&2.8\\ 
\hline 
Apr. 06, 2010 & 03:07/03:12 &0.2 (ND Long)&150 &8  &-50.5 to -48.4 & 1.07 &25.1&6.4\\ 
Apr. 06, 2010 & 03:13/05:05   &0.2 & 150&175  & -47.9 to 35.1&1.07 to 1.05 &37.9&4.9 \\
Apr. 06, 2010 & 05:06/05:11 &0.2 (ND Long)&150 &8  &35.8 to 38.8  & 1.05 &45.9&4.6 \\ 
%\hline 
%Apr. 09/10, 2010 & 0.15 & 150 & 160 &62.1/82.1 &1.29/1.63&20.4& 7.4\\ 
     \hline
Jun. 07/08, 2011 & 00:28/02:00 & 0.1              & 500 & 32 & 13/64 & 1.03/1.15 & - & 3.2\\
Jun. 07/08, 2011 & 00:44/02:12 & 0.1 (Calibrator) & 500 & 32 & 19/67 & 1.04/1.17 & - & 3.2\\
\hline
   \end{tabular}
  \end{center}
\label{logobs}
\end{table*} 

The precise detector plate scale as well as detector orientation were measured on an $\theta $ 1 Ori C field observed during the same run, and with HST (\cite{maccaughrean94} (with the same set of stars TCC058, 057, 054, 034 and 026). We found a platescale of 27.10\pm 0.04 mas per pixel and the North orientation offset was measured to be -0.26 \pm 0.09 \deg if we do not consider a systematics in the North position, or -0.26 \pm 0.3 \deg otherwise (see \cite{lagrange11} for a detailed discussion on the absolute uncertainty on the detector orientation). 

\subsubsection{ADI data}
The principle of ADI imaging is given in \cite{marois06} (see also \cite{lafreniere07}). Here, a typical ADI sequence consisted in getting  sets of saturated images (datacubes of NDIT images) at different positions on the detector, followed and precedented by a series of un-saturated PSF images recorded with a neutral density filter (ND Long).  These unsaturated images are used to get an estimate of the PSF shape for calibration purposes (photometry, shape), and fake planet simulation. On April 5/6, a few tests were made with different offsets patterns (star centered on either 
2 or 4 positions on the detector) so as to test the impact of HR4796B (which rotates on the detector during the ADI observations) on the final image quality\footnote{ As a consequence, in some offset configurations, HR4796B was out of the detector FoV, hence the number of usefull data on HR4796B is smaller than that available on HR4796A.}. 
This is important as the field of view (FoV) rotation was fast. On April 6/7, the saturated images were recorded with two offsets corresponding to two opposite quadrants on the detector. 
Both nights, the atmospheric conditions were good on average, but variable, and the amplitude of field rotation was larger than 80\deg (see Table~\ref{logobs}). The comparison between the PSFs taken prior and after the saturated images does not reveal strong variations. 

\subsubsection{SAM data}
Sparse aperture masking is obtained on NaCo by insertion of a mask in
the cold pupil wheel of the camera (\cite{lacour11}). The mask acts as
a Fizeau interferometer. It forms in the focal plane of the camera
interference fringes which are used to recover the complex
visibilities of the astronomical object. Of the four available masks
(\cite{tuthill10}), we used the 7 holes mask which gives the
highest throughput (16\%).  It is made of 1.2 meters wide circular
apertures (scaled on M1) positioned in a non-redundant fashion over
the pupil.  Minimum and maximum baseline lengths are respectively 1.8 and 6.5
meters. Each mask offered by SAM can be used in addition to almost
all the spectral filters offered by Conica.

The principle of SAM is based on its ability to facilitate the
deconvolution of phase aberrations. Phase errors are introduced by i)
atmospheric residuals and ii) instrumental aberrations (also called
non-common path errors). We used integration times of the order of the
typical coherence time of the phase errors. It permits a partial
deconvolution of the remaining atmospheric perturbation not corrected
by the AO. But most importantly, it gives  an excellent  correction of the
slowly changing instrumental aberrations. This later point is the
important factor which makes aperture masking competitive with respect
to full aperture AO.

In practice, L' SAM data on HR4796 were obtained  on June 2011. The adopted DIT was 100 ms, equivalent to a few $\tau_0$ in
the L' band. Each set of observation consists in 8 ditherings of
the telescope to produce 8 datacubes of 500 frames on 8 different
positions on the detector.  Each dither moves the star by 6 arcsec in X or in Y on the windowed
detector (512 by 512 pixels, equivalent to 14 arcsec on sky). After 8 dithers, the telescope is offset to
the K giant star HD110036 for calibration, where the very same
observation template is repeated. Four star-calibrator pairs were
obtained totalizing 64 datacubes, requiring a total observation time
of 2 hours (including overheads). Over this time, the object has
rotated by 50 degrees (the variation of the parallactic
angle).

\subsection{Data reduction}

\subsubsection{ADI data}
 Each individual ADI image was bad pixel-corrected and flat-fielded as usual. Background subtraction was made for each cube using the closest  data cubes with the star at a different offset. Data selection was also made, within each data cube and also for each data cube. Recentering of the images was done using the offsets measured by Moffat fitting of the saturated PSF. 
The data cubes were then stacked (averaged) and then reduced with different procedures that are described in details in \cite{lagrange11} and reference there-in: cADI, sADI, rADI and LOCI. These procedures differ in the way the star halo is estimated and subtracted. We recall here the differences between these various procedures, as well as new ones developed to limit the disk self-subtraction in cADI and/or LOCI: 
\begin{itemize}
\item[-] In cADI, the PSF is taken as the mean or median of all individual recentered ADI saturated images.
\item[-] To remove as much as possible the contribution of the disk from the PSF in the cADI images, we tested two slightly modified cADI reductions. In the first one, we start as usual, i.e. 
build a PSF from the median of all data, subtract this PSF to all data and rotate back the obtained residual images to align their FoV. The data are then combined (median) to get a first image of the disk. Then, to remove the disk contribution to the PSF, we rotate the disk image back to the n different FoV orientations corresponding to those of the initial images 
and subtract the median of these rotated disk images to the PSF. We obtain thus a PSF corrected (to first order) from the disk contribution. This disk-corrected PSF is then subtracted to the individual initial images; the individual residuals are then rotated back to be aligned and stacked (median) to get a new disk image (corresponding to one iteration). This ADI procedure is referred to as cADI-disk. In the second one, we  
mask the disk region in each file when used to compute the reference\footnote{In practice, we use a synthetic image (without noise) of the disk as described below, and scaled to the observed disk flux, and we define the binary mask by ascribing 0 to the pixels in the disk region that are above a given threshold (20$\%$ of the maximum flux). Then the mask is applied to each frame in the cube for the PSF estimation, so that the pixels corresponding to the disk regions in each frame are not taken into account.}. We will call this method mcADI. This method will be described in details in a forthcoming paper  (Milli et al, 2012, in prep).
%\item[-] sADI computes "PSFs" for each individual image, using a given number of images taken before or after the considered image, with a field rotation larger than a given angle expressed in FWHM at a given separation (the angle is then constant for all separations). 
\item[-] The rADI procedure (identical to \cite{marois06} ADI) selects for each frame a given number of images that were recorded at parallactic angles separated by a given value in FWHM (the same value in FWHM for each separation), to build a PSF to be subtracted from the image considered. 
\item[-] In the LOCI approach, for each given image and at each given location in the image, we compute "parts" of "PSFs", using linear combinations of all available data, with coefficients that allow to minimize the residuals in a given portion of the image. 
\item[-] To limit the impact of the disk self-subtraction on the LOCI images, we also modified our LOCI approach, masking the disk in each file whenever the disk appears in the optimization zone (see Milli et al, 2012, in prep.). We will call this method mLOCI.
\end{itemize}

The parameters used for the rADI and LOCI procedures are the following :
\begin{itemize}
\item [-] LOCI/mLOCI $\Delta r=1.4 \times\,FWHM$ below 1.6" and 5.6  $\times\,FWHM$ beyond  (radial extent of the subtraction zones); $g=1$ or $0.5$ (radial to azimuthal width ratio), $N_A=300$; separation criteria $1. \times\,FWHM$.
%\times\,FWHM^2$ (surface of the optimization zone) 
\item[-] rADI: separation criteria: $1.5 \times\,FWHM$; number of images used to compute each "PSF" : 20, $\Delta r=1.4 \times\,FWHM$ below 1.6 arcsec and 5.6 $\times\,FWHM$ beyond  (radial extent of the psf reconstruction zones)
%\item[-] sADI: the separation criteria was XXX $\times\, FWHM$ at a separation of XXX pixels; number of images used to compute each "PSF" was between XXX and XXX. 
\end{itemize}

For comparison purposes, we also performed a zero-order reduction (hereafter referred to as "nADI") which consists in, for each image, 1) computing an azimuthal average of the image (with the star position as the center of the image); we get then a 1-D profile, 2) circularizing this 1-D profile to get a 2-D image centered on the star position, 3) subtracting the obtained image to the initial image to get corrected image. We then derotate and stack all the "corrected" images.  nADI clearly does not benefit from the pupil stabilization and is not to be considered as  a  real ADI reduction procedure, but can help in some cases disentangling artefacts produced by ADI reductions from real features.

The data obtained on the 6th and 7th were reduced separately and then averaged. As they happened to have similar S/N ratio, a simple averaging was made. 

\subsubsection{SAM data}
The first step to reduce the SAM data is to clean the frames. This can
be done in the same way as any classical imaging method in the infrared. In
practice, we flatfielded the data and subtracted the background. The
background was estimated by taking the median value of the 8 datacubes
of a single observation set.

As any interferometric facility, the observable parameters of SAM are
fringes. The information lays in the contrast (which, once normalized,
is called visibility) and the phase. Contrasts and phases are obtained
by least square fitting of the diffraction pattern. Since the fitting
of sinusoidal curves is a linear least square problem, a downhill
algorithm to find the maximum likelihood was not required. Instead,
inversion was done by projection of the datacubes on a parameter space
defined by each complex visibility fringes. The matrix used for
projection is determined by singular value decomposition of a model of
the fringes. In the end, we checked that it gives exactly the same
result as a least square minimization algorithm of the kind of
conjugate gradient (but much faster).  The fringes are modeled by cosines of given frequency multiplied by the PSF of the Airy pattern of a single hole. Wavelength bandwith is accounted for by smearing the pattern over the filter bandpass.

As a result, we get a single complex value for each baseline and each
frame. 
 They are used to compute the bispectrum, which is summed over
the 8 datacubes which correspond to a single acquisition. Then, the
closure phases are obtained by taking the argument of the
bispectrum. One set of closure phase is obtained for an observation set
which takes around 8 minutes. Over that time, the parrallactic angle
changes less than 6 degrees, which effect is neglected (baselines
rotation during an observation set is not accounted for). The final step consists in calibrating the closure phase of HR4796 by
 subtracting the corresponding values obtained for the red giant
(HD110036).

\section{Data simulations}
Obviously, ADI  affects the resulting disk shape because of disk self-subtraction. This effect is expected to be more important as the disk inclination with respect to line-of-sight decreases. Also, the different ADI reduction procedures will impact differently the disk shape. A general study of the impact of ADI on disk reductions will be presented in a forthcoming paper (Milli et al, 2012, in prep.). In this paper, we concentrate on the HR4796 case and we monitor this impact using fake disks, as done in \cite{lagrange11}. 
\subsection{Assumptions}
To simulate the HR4796 disk, we assumed, following \cite{jca99} a radial midplane number density  distribution of grains $\propto$ ((r/r${_0}$)$^{-2\alpha_{in}}$ + (r/r${_0}$)$^{-2\alpha_{out}}$)$^{-0.5}$. We chose r${_0}$ = 77.5 AU, $\alpha_{in}$ =35 to ensure a very sharp inner edge, and $\alpha_{out}$ = -10,  as assumed by \cite{thalmann11}. The vertical distribution is given by: 
$$ Z(r,z) = e^{-\left(\frac{|z|}{\xi}\right)^\gamma}$$ where the height scale 
$\xi=\xi_0 \left(\frac{r}{r_0} \right)^\beta$ is 1 AU at 77.5 AU. The disk flaring  coefficient is $\beta$=1 and the coefficient $\gamma=2$ ensures a gaussian vertical profile. The disk is inclined by 76 degrees (a pole-on disk would have an inclination of 0 degree), and we assumed an isotropic scattering (g=0), as \cite{hinkley09} polarimetric measurements indicate a low value for g (0.1-0.27). The disk was simulated using the GRaTer code (\cite{jca99}; \cite{lebreton11}). It will be referred to as HR4796SD. The ring FWHM thus obtained is 0.14" (before reduction) under such hypothesis. We also considered another disk, with all parameters identical to those of HR4796SD, but with $\alpha_{out}$ = -4; this disk (referred to as HR4796blowoutSD) is representative of the outer density distribution that would be observed if the outer brightness distribution was dominated by grains expelled by radiation pressure as in the case of \bp (\cite{jca01}).
 
\subsection{Simulated disk images}
 The flux of the simulated disk is scaled so as to have the same number of ADU (at the NE ansae) as in the real disk, once both simulated and real data are reduced by cADI. When brighter disk are needed, a simple scaling factor is applied.  The simulated projected disks are then injected in a datacube, at each parallactic angle, corresponding to each real data file, and are then convolved either by a theoretical PSF matching the telescope and instrument response, or the average of the real PSFs taken prior to and after the saturated images. Each image is added to each frame of the original data cube, with a 130\deg or 90\deg  offset in PA with respect to the real disk, so as to minimize the overlap between both disks. The datacubes are then processed by nADI, CADI, mcADI, LOCI and mLOCI.

\section{The HR4796A disk}
\subsection{Disk images: qualitative view}
\begin{figure}
\centering
%\includegraphics[angle=0,width=0.9\hsize]{figures/image_combination_grey_wo_sadi.ps}
%\includegraphics[angle=0,width=0.9\hsize]{figures/image_disk.ps}
%combination_nadi_cadi_cadids_radi_loci.ps}
%\includegraphics[angle=0,width=0.9\hsize]{figures/combination_nadi_cadi_cadids_mcadi_radi_loci_mloci.ps}
\includegraphics[angle=0,width=0.9\hsize]{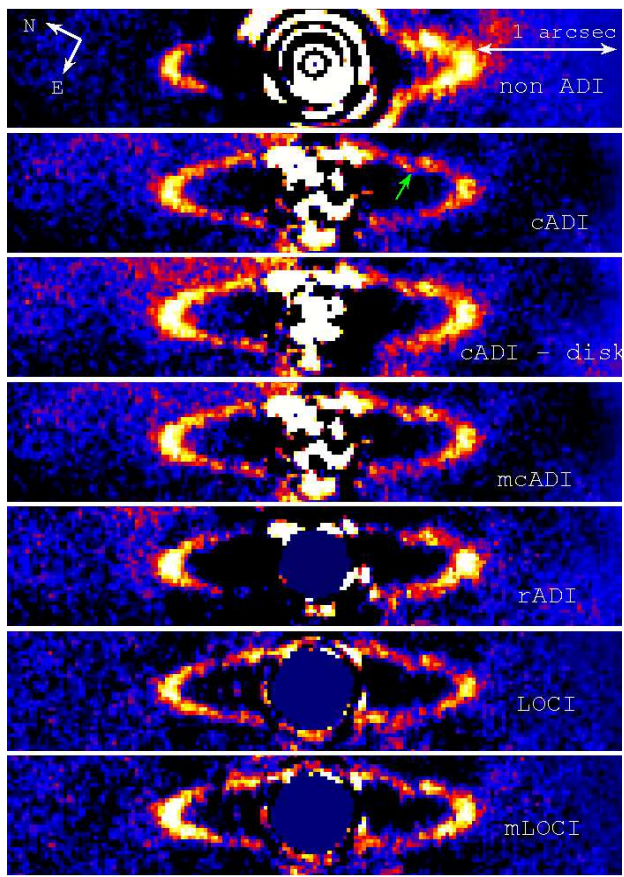}
 \caption{HR4796A disk at L' (linear scale). The pixel scale is 27 mas/pixel. From top to bottom, the same data reduced with  nADI, cADI, cADI-disk, mcADI, rADI, LOCI, and mLOCI (see text). Note that the color codes are identical for all cADI reductions reductions on the one hand, and LOCI reductions on the other hand, to enable comparisons within a given method, but different cuts are used for cADI, rADI and LOCI reductions.}
\label{planche_disk}
\end{figure}

Figure~\ref{planche_disk} shows the images obtained when combining the data obtained on the 6th and 7th of April. Images resulting from the ADI reductions described above are showed: cADI, cADI-disk, mCADI, rADI, LOCI, and mLOCI. We also show for comparison the image resulting from nADI reduction. 

%\vfill\eject

The disk is clearly detected at "large" separations from the star with the nADI reduction, and is, expectedly, lost in the Airy rings closer to the star. The parts closer to the star in projection are revealed only by the real ADI reductions, and actually, the disk is more completely detected than in previously published images, in particular the west side is almost continuously detected. The masking greatly improves the image quality of the LOCI image; the impact of mcADI with respect to cADI is, expectedly, less important, even though the flux restitution is increased. Nevertheless, the dynamical range of our images is lower than that of the recently AO  published images. This is because the present data are obtained at L', with a higher Strehl ratio, whereas the previous ones were obtained at shorter wavelengths, with lower Strehl ratio, but with detectors that have much lower background levels. 

The ring appears very narrow in our L'-data, barely resolved: the FWHM measured on the PSF is 4.1 pixels (0.11"), while the ring FWHM is $\simeq$ 5.7 pixels (0.15") (NE) and resp. 5.0 pixels (0.13") (SW) on the cADI data. We will however see below that the ADI reduction has an impact on the observed width.
%This is in contrast with the data obtained by \cite{thalmann11} at H band, a wavelength 2.5 times smaller, hence providing a spatial resolution at least twice better.

 %We see in Figure~\ref{planche_disk} visual differences in disks shapes between e.g. the cADI and LOCI images. The LOCI disk seems more triangular than the cADI one close to the semi-major axis, and with an enhanced brightness close to the semimajor axis than elsewhere in the disk. Such effects are artefacts from the reduction procedure (see the various LOCI simulations of bright fake disks throughout this paper; see also Milli et al, 2012, in prep.). 

\begin{figure*}
\centering
%%\includegraphics[angle=0,width=0.9\hsize]{figures/cadi_loci_thalmann_v2.ps}
%image_combination_fakedisk_thalmann_color_wo_radi_sadi.ps}
%image_combination_fakedisk_thalmann_color_wo_sadi.bck.ps}%% %
\includegraphics[angle=0,width=0.4\hsize]{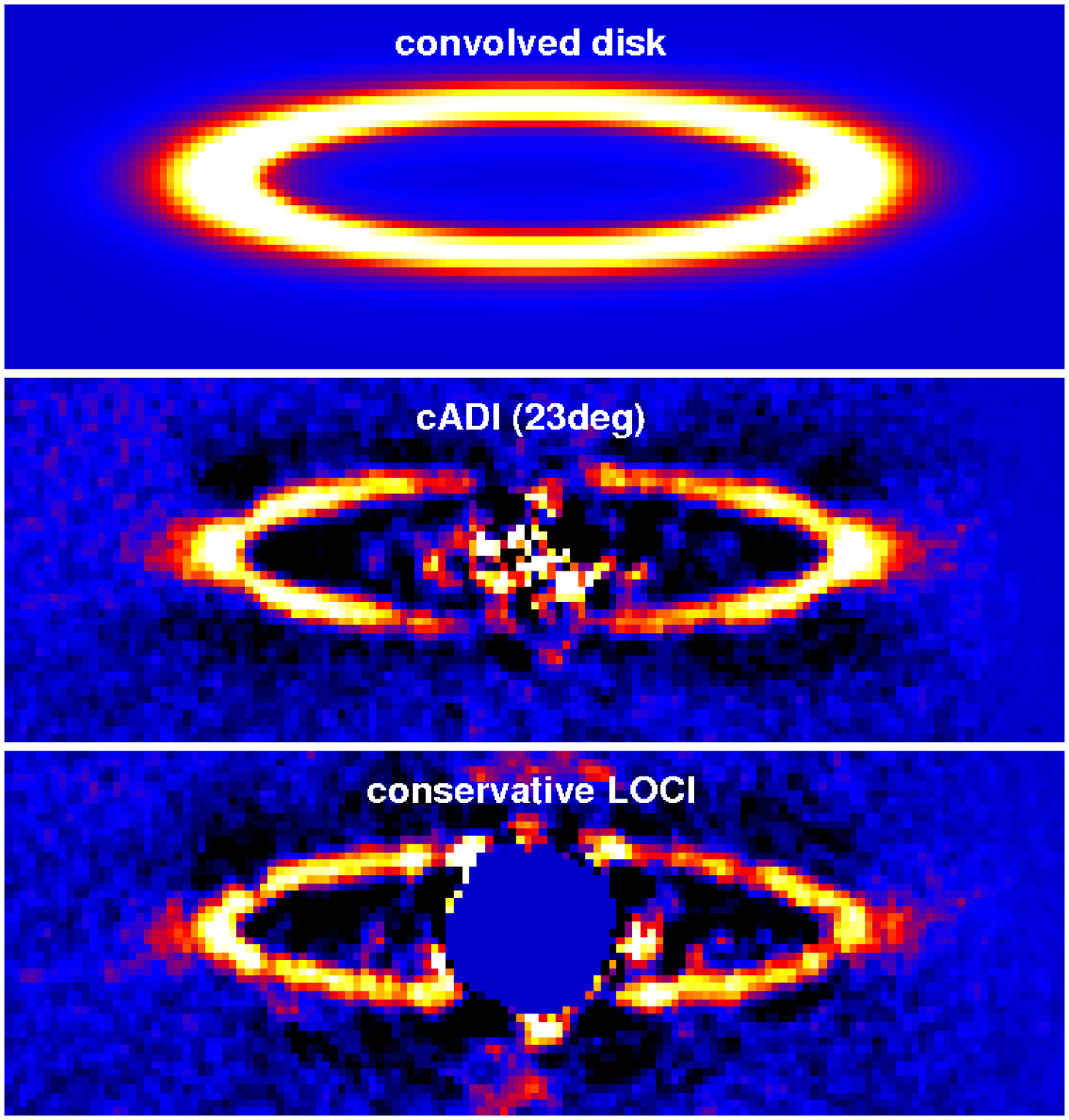}
\includegraphics[angle=0,width=0.4\hsize]{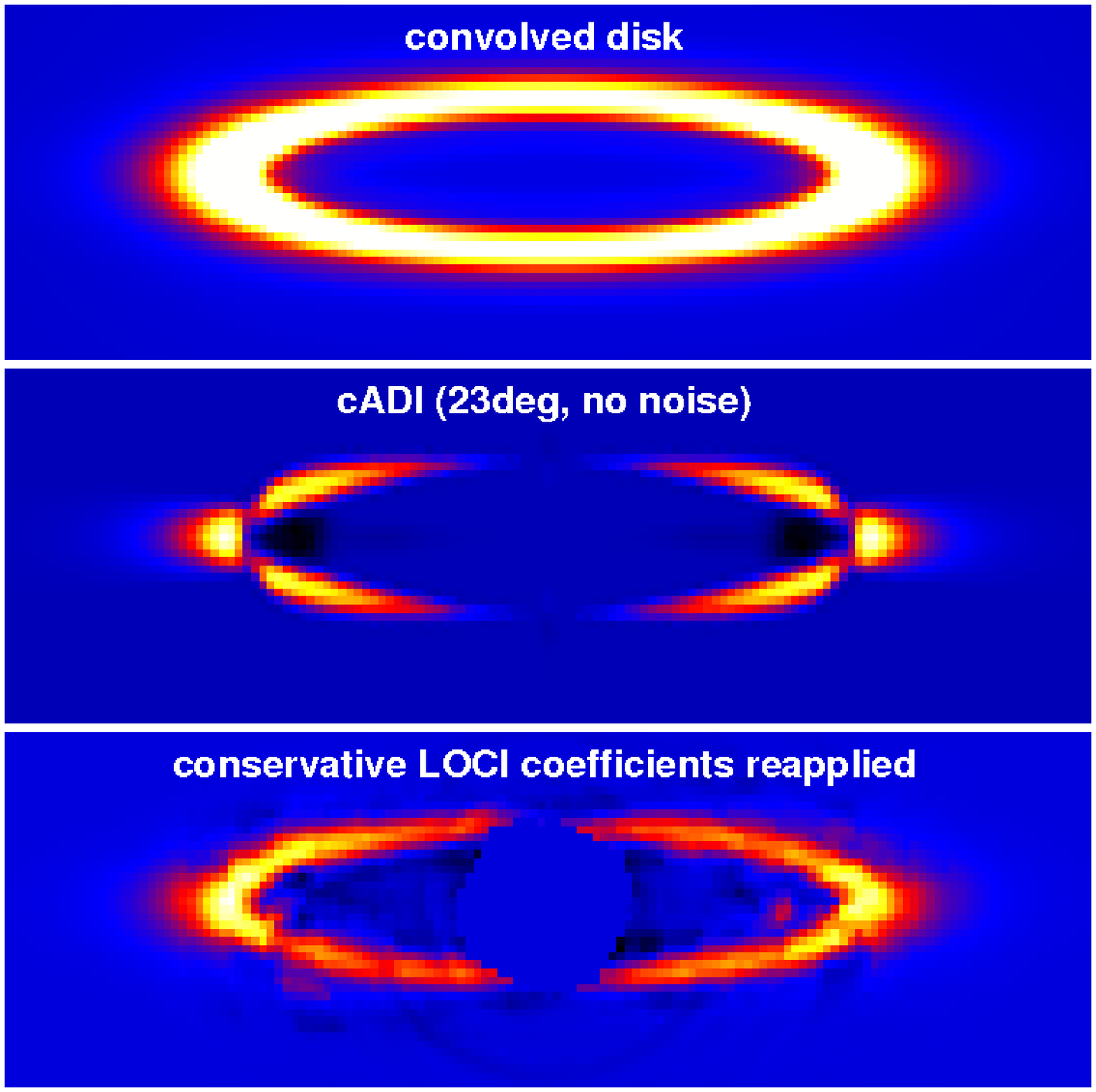}
  \caption{ Left) From top to bottom 1) simulated HR4796SD-like disk (projected, no noise); 2) 
simulated cADI reduction of this disk once inserted into our data cube, and convolved by a theoretical PSF. We assumed a 23 degrees FoV rotation as Thalmann et al (2011). 3) idem with LOCI reduction. Right) Idem without noise.}
\label{planche_simus_streamers}
\end{figure*}

\begin{figure*}
\centering
\includegraphics[angle=0,width=0.9\hsize]{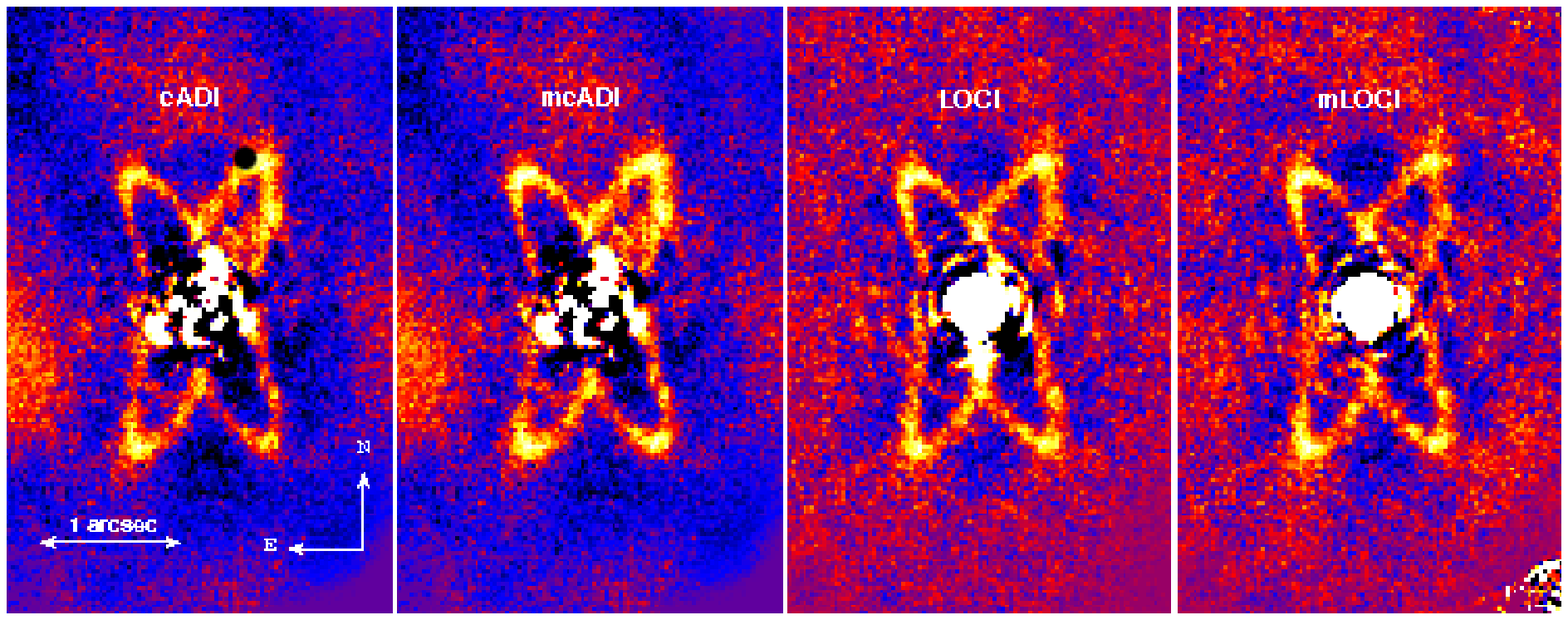}\\
\includegraphics[angle=0,width=0.9\hsize]{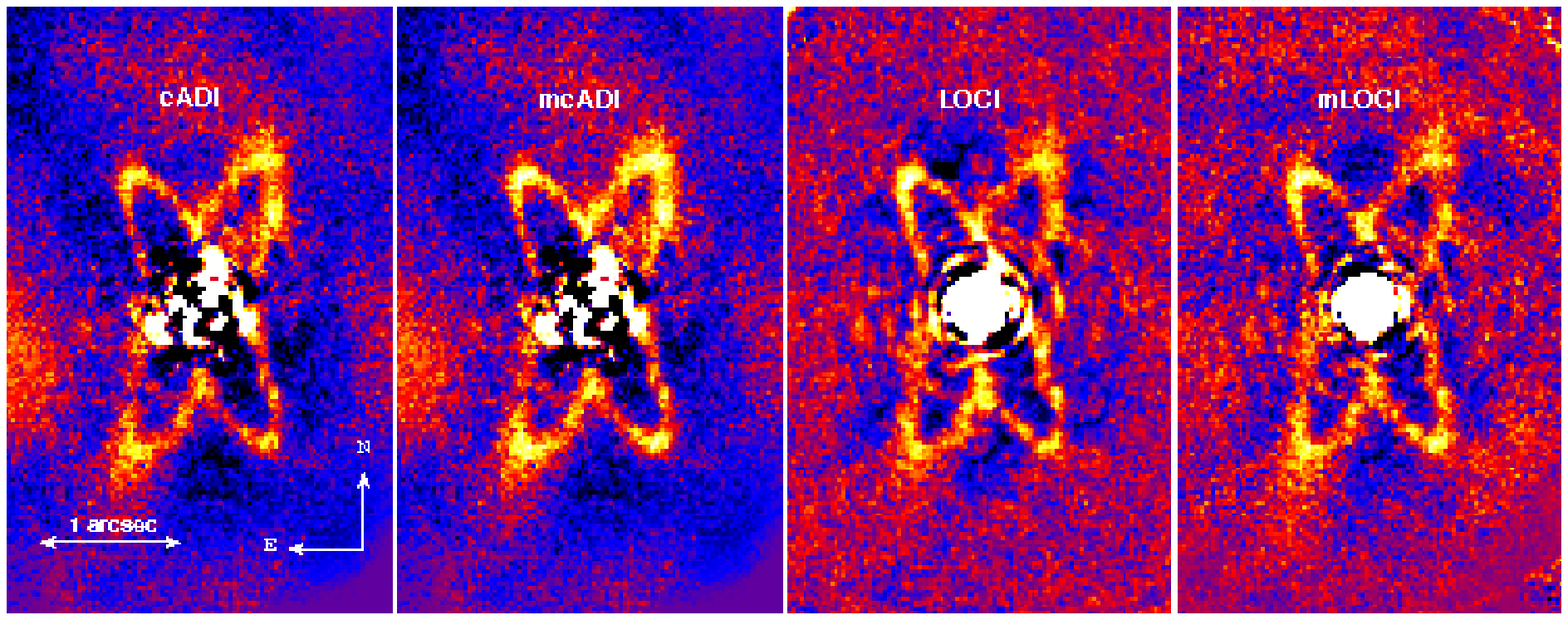}\\
%combination_cadi_loci_fakediskinjected_blowout.ps}
%image_combination_wo_sadi_simus.ps}
%fait a partir de image_combination_color_fakedisk_wo_sadi.jp.jpg
%combination_cadi_loci_fakediskinjected.ps}\\
  \caption{Top: simulated disk (model HR4796SD) inserted into the data: images (linear scale) reduced with cADI, mcADI, LOCI and mLOCI, for a rotation angle amplitude of 85 deg similar to that of our data.  Bottom: idem for a disk model assuming $\alpha$$_{out}$ =-4 (model HR4796blowoutSD).  North is up and East is to the left. The real disk is NE-SW oriented; the simulated one is NW-SE oriented.}
\label{planche_disksimu_noise}
\end{figure*}

 \cite{thalmann11} report a relative enhancement of the disk brightness in the outer part of the ring, along the semi-major axis, that they describe as  streamers emerging from the ansae of the HR 4796 disk. Our images do not show such strong features. To test whether such features could be due to the ADI reduction and/or data characteristics, we built cADI and LOCI images of a bright\footnote{we assumed a disk 10 times brighter than the actual one so as to take into account the better SN of \cite{thalmann11} H data, compared to the present L' ones.} HR4796SD-like disk, inserted into our data cube, convolved by a theoretical PSF and, in the case of LOCI, reduced with Thalmann et al. LOCI parameters, assuming a 23 degrees FoV rotation. The input disk images are shown in  Fig.\ref{planche_simus_streamers}, as well as the reduced ones. The reduced images clearly show features similar to those observed by \cite{thalmann11} in their Figures 1 and 3. We also show in Fig.\ref{planche_simus_streamers} the same simulations of a similar disk inserted into a data cube matching the parallactic angle excursions, but assuming no noise. Note that for the LOCI reduction of noiseless data, we took the coefficients derived from the previous LOCI reduction (taking noise into account); this is necessary to avoid major artefacts when using LOCI on noiseless data.  
 The later simulations (no noise) highlight the reduction artefacts. Other simulations are provided in Milli et al. (2012, in prep.). Hence, we conclude that the features indicated as "streamers" are in fact artefacts due to the data characteristics (FoV rotation amplitude, number of data, SN) and data reduction. The fact that we do not see them in the present data is due to our relatively lower dynamical range, and to the larger FoV  rotation amplitude. To check this, we injected  a fake disk (HR4796SD), with a flux similar to the observed one in the actual datacubes (with a 130\deg offset in PA to avoid an overlap of the disks). We processed the new data cubes as described above with cADI, mcADI, LOCI and mLOCI. The resulting images are shown in Figure~\ref{planche_disksimu_noise}; the artefacts are not detectable. This is also true when considering a fake disk with blowout (HR4796blowoutSD).

\begin{figure}
\centering
\includegraphics[angle=0,width=0.9\hsize]{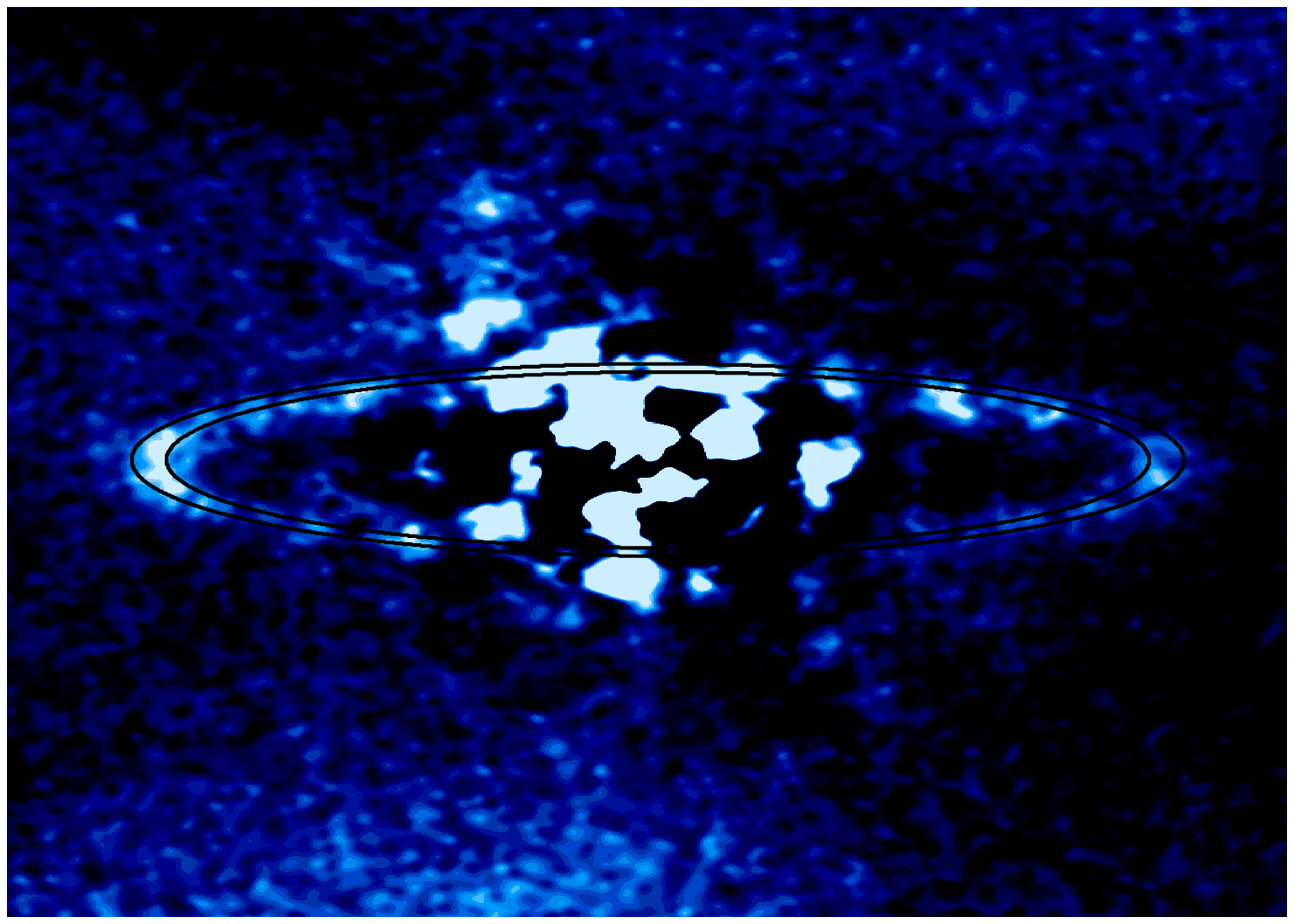}\\
\includegraphics[angle=0,width=0.9\hsize]{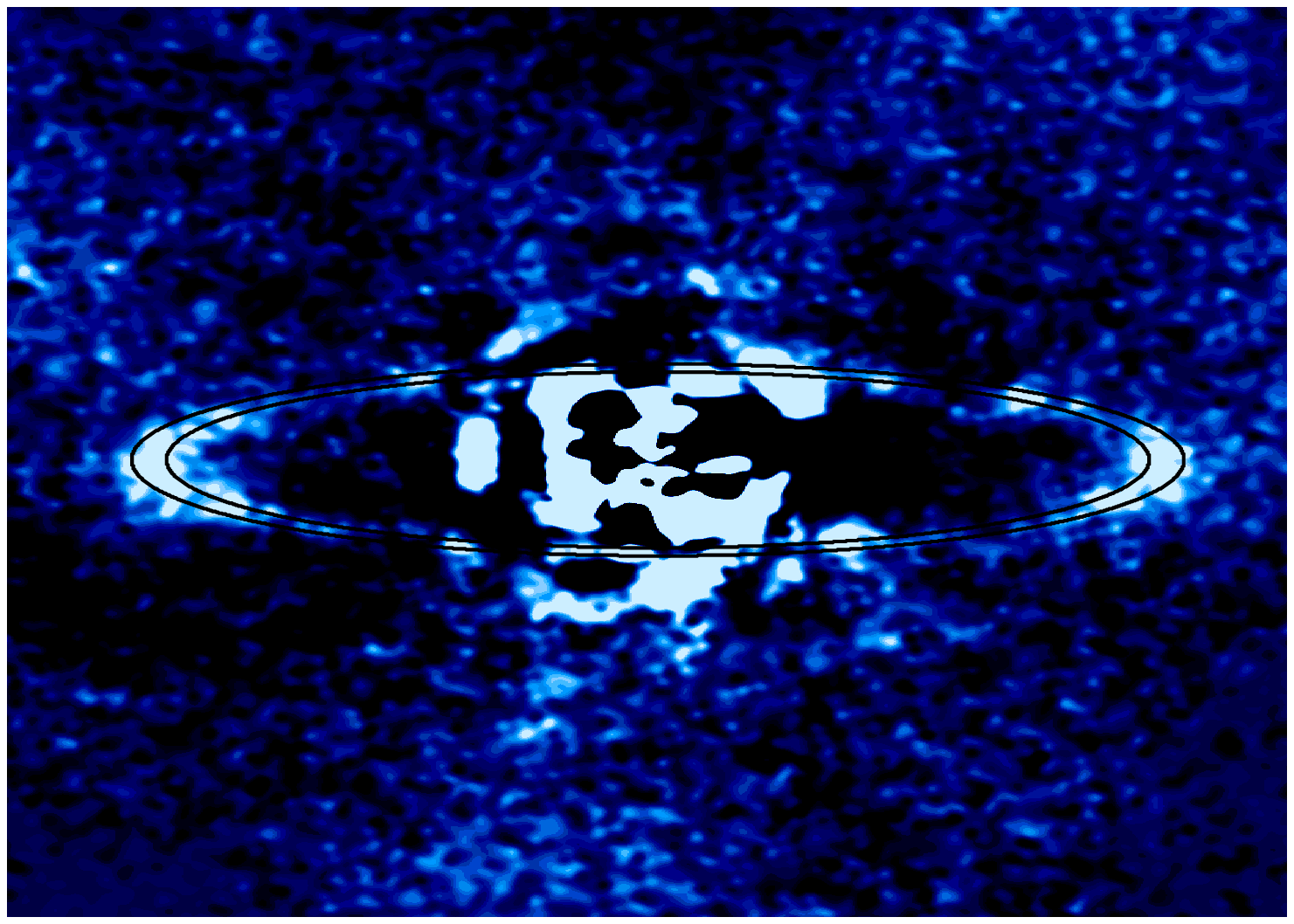}\\
%combination_cadi_loci_fakediskinjected_blowout.ps}
%image_combination_wo_sadi_simus.ps}
%fait a partir de image_combination_color_fakedisk_wo_sadi.jp.jpg
%combination_cadi_loci_fakediskinjected.ps}\\
  \caption{Top: cADI image of HR4796 on April, 6th. Bottom: idem on April, 7th.}
\label{planche_disk_6and7}
\end{figure}

\begin{figure}
\centering
\includegraphics[angle=0,width=0.9\hsize]{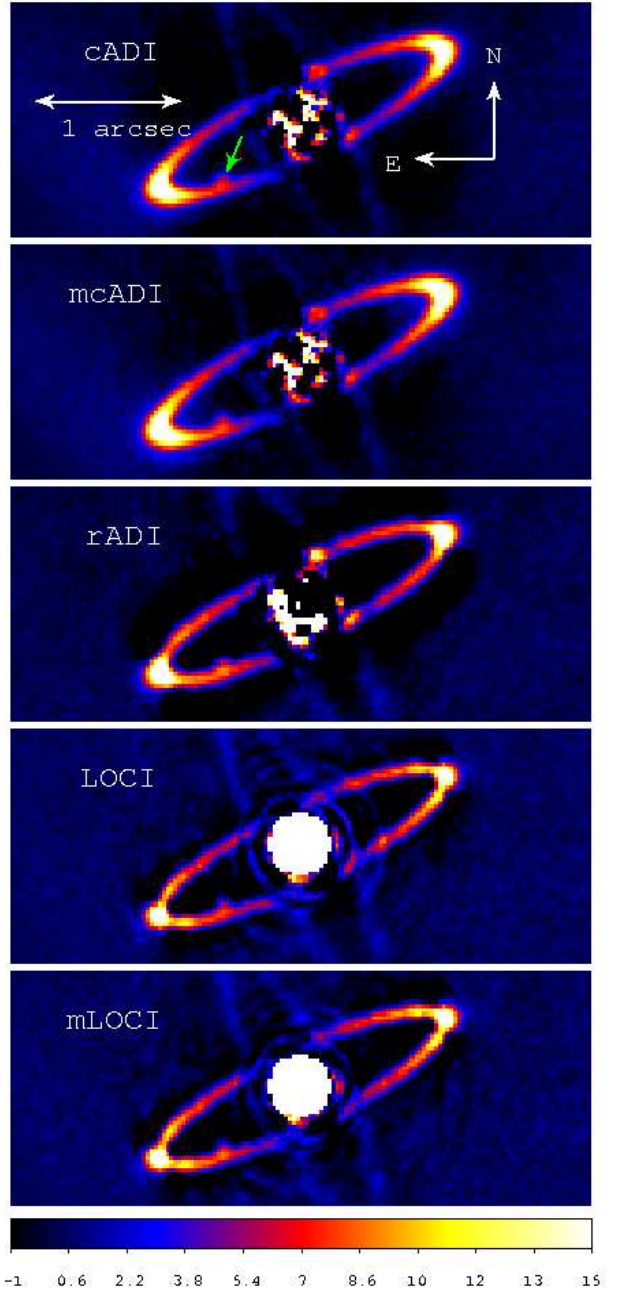}
  \caption{ cADI, mcADI, rADI, LOCI and mLOCI images of simulations of fake disk + fake point-like source (indicated by a green arrow). }
\label{planche_diskplanetsimu_noise}
\end{figure}

Finally, we note a small distorsion in the SW disk towards the inner region of the cADI and LOCI images, at (r,PA) between (19 pix, 235\deg) and (28 pix, 220\deg). The feature, indicated by a green arrow in Figure~\ref{planche_disk}, is however close to the noise level. If we go back to the individual images taken on April, 6th and April, 7th, we see that this feature is barely detectable on the April 6th eventhough in both cases, there is a very faint signal inside the disk ellipse (see Figure~\ref{planche_disk_6and7} for the cADI images). Hence, in the present data, this feature could be due to noise. However, it seems to be at the same position as that pointed by \cite{thalmann11} in their data as well as \cite{schneider09} data as well as, in L'-data obtained at Keck by C. Marois and B. McIntosh (priv. com.). 
 In \cite{thalmann11} data, it appears as a loss of flux in the annulus. In the L' data, we see rather a distorsion in the disk and a possible very faint additional signal at the inner edge of the disk. However, the ring does not appear as azimuthally smooth in the \cite{thalmann11} or our data, due to ADI reduction and limited SN, so it is not excluded at this stage that this feature might be an artefact. Clearly, new data are needed to confirm this structure, and if confirmed, to study its shape as a function of wavelength.  If confirmed, its origin should be addressed. In the context of the HR4796 system, an interesting origin to be considered is the presence of a planet close to the inner edge of the disk. To test the impact of the ADI reduction procedure on a disk + close planet system at L', we inserted a fake point-like source close to our fake disk HR4796SD (rotated by 130\deg with respect to the real disk, convolved by the observed PSF, and inserted in the datacube) inner edge, and processed the data as described previously. For these simulations, we assumed a disk about 10 times brighter than the real one. We run several simulations with different values of planet fluxes and distances to the disk.  For some values of the planet position and flux, we were able to reproduce a disk appearent distorsion, especially in the LOCI images. A representative example is given in Figure~\ref{planche_diskplanetsimu_noise}; in this case, the planet flux would correspond to a 2 MJup mass for a disk brightness similar to the HR4796 one. 

%Using the derived flux loss for the fake planet and the observed residual flux after ADI reduction, we conclude that if the point-like signal in the real data was due to a planet, this planet would have a mass of about 2 MJup, using the COND03 models (\cite{baraffe03}) and assuming an age of 8 Myr for the system. 

%We also show in Figure~\ref{planche_disksimu_noise} the images corresponding to a disk with $\alpha$=-4. 

\subsection{Disk properties}
\subsubsection{Disk geometry}
\begin{figure}
\centering
\includegraphics[angle=0,width=0.95\hsize]{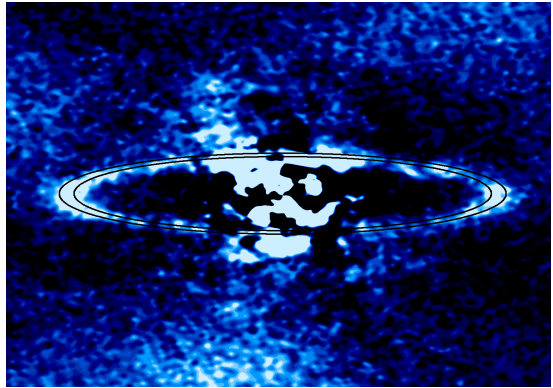}
%plot_map_cadi_ellipse.jpeg
  \caption{Best elliptical fit to the L' cADI data. NE is to the left and SW to the right.}
\label{fit_ellipse}
\end{figure}

We fitted the observed disk by an ellipse, using the maximum regional merit method, as in \cite{buenzli10} and in \cite{thalmann11}. The resulting semi-major axis a, semi-minor axis b, disk center position along the semi-major axis (xc), and the semi-minor axis (yc), and inclination (defined as arccos(b/a)) are provided in  Table~\ref{ellipse}, and the fit is showed in Figure~\ref{fit_ellipse}. These parameters are derived from the selection of the best fits, defined as those with parameters within 5$\%$ of the best fit (best merit coefficient). The uncertainties associated to these measurments take into account the dispersion within this 5$\%$ range, and the other sources of uncertainties that are described hereafter. 

To estimate the impact of the PSF convolution and ADI process on the ellipse parameters, we used our simulated disk HR4796SD (without noise) and fitted the disk with an ellipse before and after the PSF convolution and the ADI reduction. For (a,b), differences of (-0.03;0.12) pixels were found with cADI, and (-0.06;-0.3) pixels with LOCI. For (xc, yc), no significant differences were found with cADI and very small with LOCI. No significant difference was measured on the inclination with cADI while a difference of -0.4\deg was found with LOCI. Finally, no significant differences were found on the PA. We corrected the measured values on the disk from these biases. We also inserted a model disk HR4796SD in the data cube (at 90 degrees), and processed the data. The differences found between the parameters of the injected disk and the ones of the recovered disk are compatible with the ones obtained in the case "witout noise". 
%These biases were taken into account in the values given in Table~\ref{ellipse}. 

Besides, the imperfect knowledge of the PSF center may also affect the results. To estimate this impact quantitatively, we first estimated the error associated to the PSF center, as in \cite{lagrange11}. The error was found to be [0.,0.27] pixel on the x-axis and [-0.06, +0.04] pixel on the y-axis of the detector. It appears that this imperfect knowledge on the PSF center does not significantly affect the values of (a,b). It impacts the uncertainty of the ellipse center by up to 0.2 pixel along the major and minor axis, and the disk PA (0.24 \deg ). The uncertainty on the PA measurement is found to be 0.15\deg in cADI.

Finally, the PA measurement is also impacted by the uncertainty on the true North Position (0.3\deg ; see a discussion related to this last point in \cite{lagrange11}). 
 
%The results are given in the second part of Table~\ref{ellipse}. In this part, the first line gives the parameters resulting from the fit of the disk, before any ADI procedure is applied, and the other lines provide the results from the fit after ADI is applied. We see from the measurements of the simulated disks that the cADI process does not affect the disk parameters at least with the present precisions, where as LOCI slightly affects the measured ellipse semi-minor axis (-0.5 pixel) and, consequently, the disk inclination . effet du blow out???

 \begin{table*}[t!]
  \caption{Ellipse parameters of the observed disk: semi-major axis (a, mas), semi-minor axis (b, mas), position of the center of the ellipse with respect to the star ((xc, yc), expressed in mas), disk PA (deg), and inclination (i, \deg). }
%The first part corresponds to the actual measurements; the second part to a simulated disk with $alpha$ =-10 (ie a steep outward density distribution), while the third one corresponds to a disk with $\alpha$ = -4, corresponding to an ouward density distribution representative of dust expelled by radiation pressure.  
\label{stats}
\begin{center}
\begin{tabular}{ l l l l l l l  }
 \hline
measured & a (mas)& b (mas) & xc (mas)& yc (mas)& PA (deg) & i (deg)\cr
\hline 
%cADI &39.72\pm 0.2 	&10.1\pm	0.09&	0.85\pm	0.13&	0.27\pm	 0.07	&	26.85\pm 0.15&	75.27\pm 0.15	\\
%LOCI&39.4\pm	0.2&	9.5\pm	0.11&	0.8\pm	0.16&	0.4\pm	0.07&	26.6\pm	0.15&	76.05\pm	0.18\\
%cADI &39.72\pm 0.2	&10.1\pm	0.09&	0.85\pm	0.24&	0.27\pm	 0.21	&	26.85\pm 0.40&	7	\\
%{\bf cADI non corrected& 1076 .41\pm 5.4 	& 273.71\pm 2.44	& 23.05	\pm 10.84 & 	7.32\pm	 5.69	& 26.86\pm 0.28 &	75.27\pm 0.21 \\
cADI & 1077.2\pm 4.1 	& 270.5\pm 2.4	& 22.8	\pm 6.5 & 	7.6\pm	 5.7	& 26.85\pm 0.15 &	75.28\pm 0.15 \\
%LOCI non corrected& 1067.74\pm 5.42 &	257.45\pm	2.98 &	21.68\pm 7.05&	10.84\pm	5.69&	26.60\pm	0.15&	76.05\pm	0.18\\
LOCI & 1070.2\pm 5.4 &	266.9\pm	3.0 &	19.5\pm 6.9&	10.8\pm	5.7&	26.60\pm	0.32&	75.58\pm	0.3\\
%cADI & 39.8\pm 0.15 &	10.1\pm 0.09&	0.9	\pm 0.13 &0.3\pm 0.07&	26.8	\pm 0.15&75.3\pm 0.15\\
%rADI & 40.1	\pm 0.15&10.2\pm 0.14&	1.1\pm 0.14&	0.3\pm 0.1&	27	\pm 0.15&75.26\pm 0.21\\
%LOCI & 39.4	\pm 0.2 &9.5	\pm 0.11 &0.8\pm 0.16 &	0.4\pm 0.07 &26.6\pm  0.15&76.05 \pm 0.18	\\
\hline
% model & a (pix)& b (pix) & xc (pix)& yc (pix)& PA (deg) & i (deg)\cr
%\hline 
%input, $\alpha$=-10 & 40.7\pm 0.26	&10.2\pm 0.1&	0\pm 0.22&	0\pm 0.1&	26.4\pm 0.22&	75.49\pm 0.16\\
%cADI & 40.7 \pm 0.23&	10.2\pm 0.12&	-0.1\pm 0.19&	-0.1\pm 0.1&	26.4\pm 0.21&	75.49 \pm 0.19\\
%rADI & 41.1\pm 0.23&	9.7\pm 0.1 &	-0.1\pm 0.21 &	-0.1\pm 0.21&	26.4\pm 0.13 & 76.35 \pm 0.16\\
%LOCI & 40.8\pm 0.17&	9.7\pm 0.1 &	-0.1\pm 0.16 &	-0.1\pm 0.1&	26.4\pm 0.13 &	76.25\pm 0.16\\
%     \hline
%input, $\alpha$=-4 & 42.2\pm0.3&	10.7\pm 0.18 &	0\pm 0.26&	0\pm 0.14 &	26.4\pm 0.27 &	75.31\pm 0.27 \\
%cADI & 42.2\pm 0.23 &	10.7\pm 0.11&	0\pm 0.24	&0\pm 0.13&	26.4\pm 0.23&	75.31\pm 0.18\\
%rADI & 42.1\pm 0.35 &10.4\pm 0.12&	0\pm 0.27 &	0\pm 0.11&	26.4\pm 0.2&	75.7\pm \\
%LOCI & 42\pm 0.16 &	10.4\pm 0&	0.8\pm 0&	0\pm 0.15&	25.4	\pm 0&75.66\pm 0.06\\
%\hline
    \end{tabular}
  \end{center}
\label{ellipse}
\end{table*} 

Our data show an offset from the star center of $\simeq$ 22 mas on the cADI images ($\simeq$ 20 mas on the LOCI data) of the center of the fitted ellipse along the major axis, and to the South. Given the uncertainty associated to this measurement, 7 mas, we conclude that the observed offset is real. This offset along the major axis is in agreement with previous results of \cite{schneider09} (19\pm 6 mas) and \cite{thalmann11} on their LOCI images (16.9 \pm 5.1 mas). The latter detected moreover an offset of 15.8\pm 3.6 mas along the minor axis, which was not detected by \cite{schneider09}; the measured offset on our cADI images is about 8 mas \pm 6 mas; hence very close to the error bars, so the present data barely confirm the offset found by \cite{thalmann11}. %As the uncertainties associated to the \cite{thalmann11} results are not detailed, it is difficult to further compare both results.

\subsubsection{Brightness distribution}

\begin{figure*}
\centering
\includegraphics[angle=-0,width=0.45\hsize]{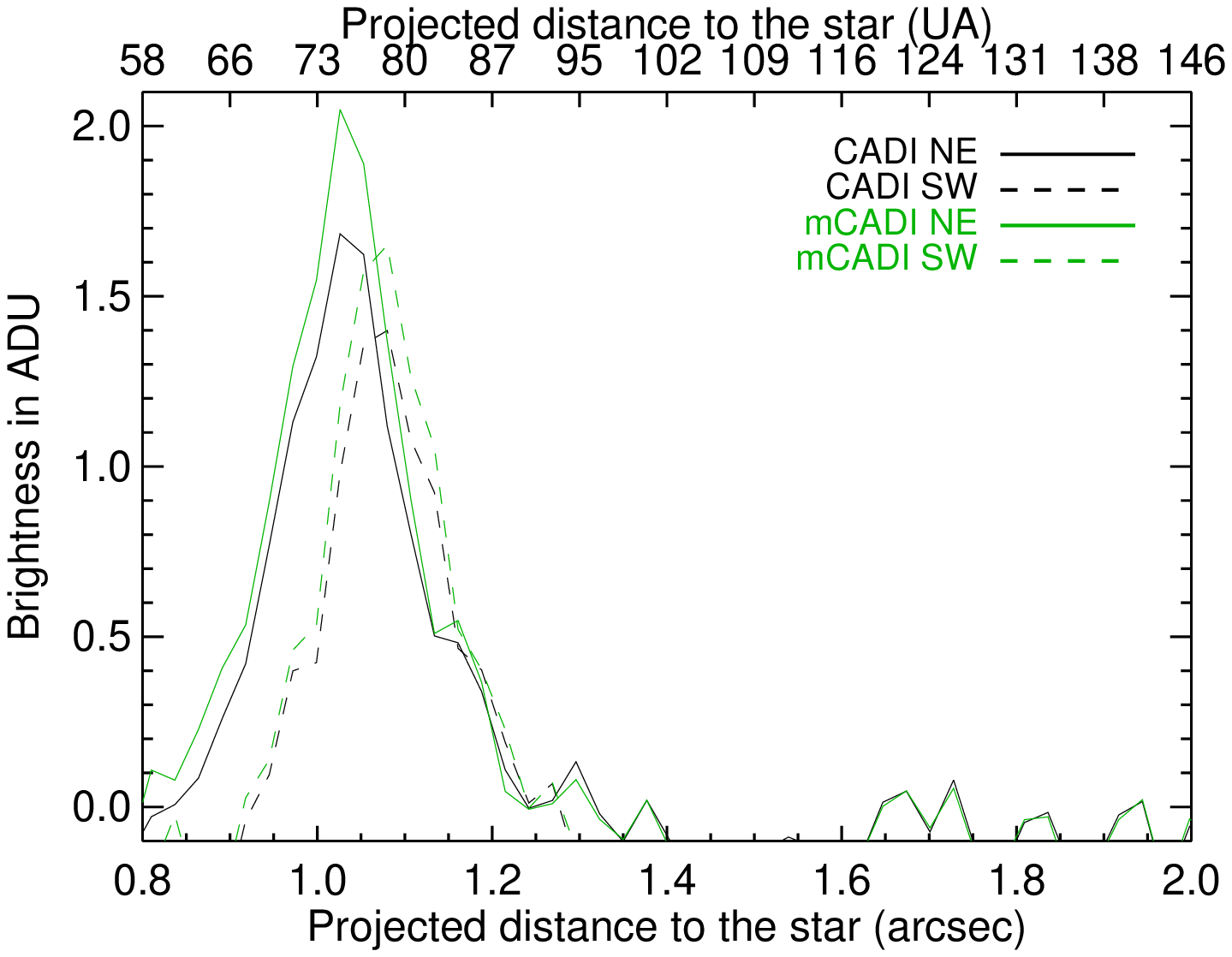}
%obs_profiles_cadi_nolog.eps}
\includegraphics[angle=-0,width=0.45\hsize]{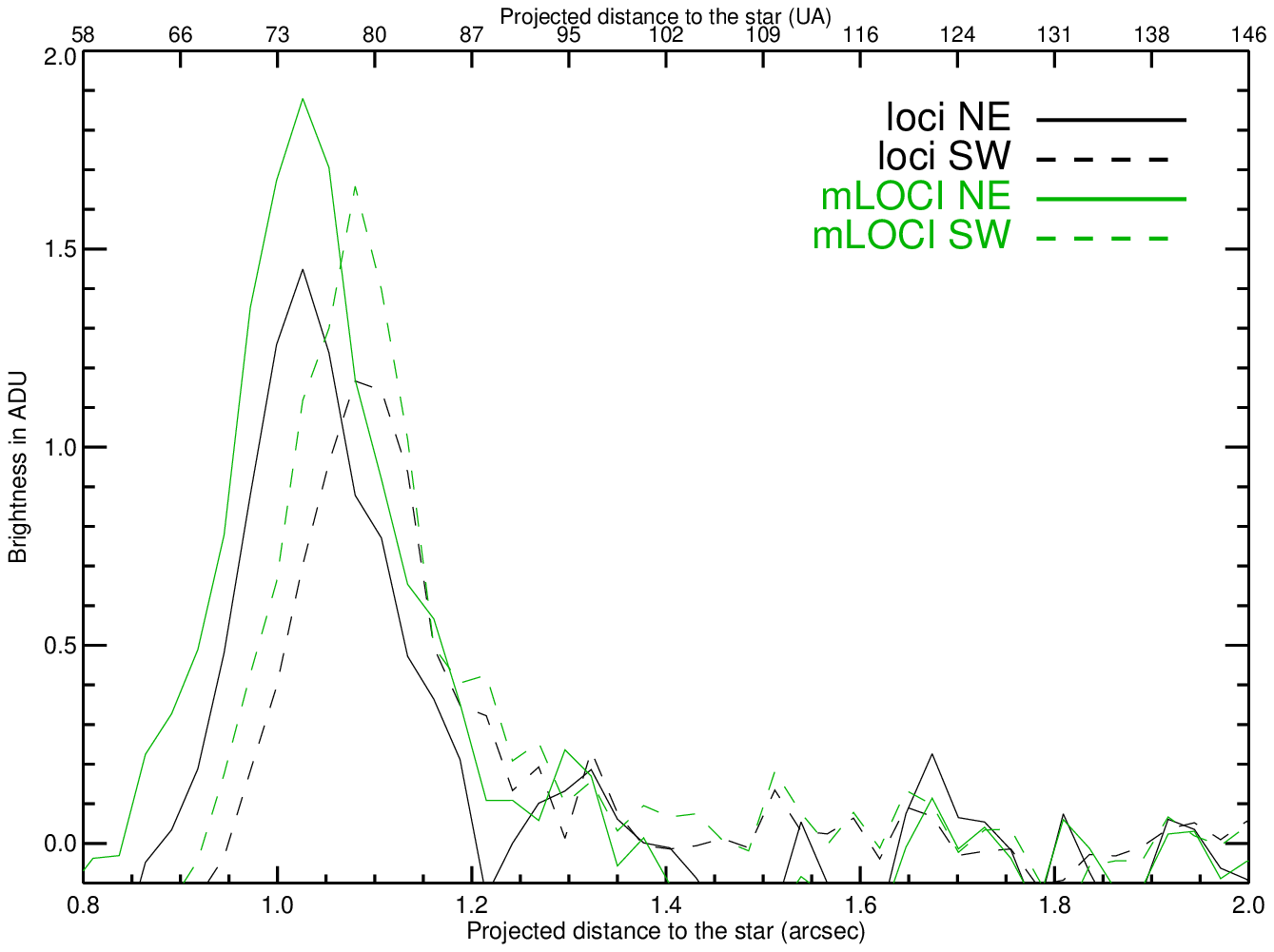}
%obs_profiles_loci_nolog.eps}
  \caption{Observed SBD for the HR4796A disk at L' along the major axis (after a binning of 5 pixels  perpendicular to the major axis): cADI and mcADI reductions (left); LOCI and mLOCI reductions (right). We note that the peaks of the SBD on the NE and SW do not coincide which is due to the shift of the ellipse center along the major axis.}
\label{obs_profiles}
\end{figure*}

Figure~\ref{obs_profiles} shows the observed radial surface brightness distributions (SBD) for the HR4796A disk at L' along the major axis after cADI, and LOCI reductions. The SBD extraction was made using a 5 pixel vertical binning. The dynamical range of our data is small (factor of 10); it is improved with masking technics, thanks to a lesser  disk  self-subtraction (see also Figure~\ref{obs_profiles}). Yet, the disk being only slightly resolved, we cannot perform meaningful slope measurements on our data. Indeed, the slope of the surface brightness distribution depends on several parameters: the PSF, the amplitude of the FoV rotation, the ADI procedure, the binning used for the extraction of the SBD, and the separation range on which the slopes are measured. In the present case, the separation range is too small to allow a proper measurement of the slope: we run simulations of the HR4796SD disk without noise and checked that indeed, measuring the slopes between the maximum flux and the threshold corresponding to the noise on the actual data gave slopes very different (much higher) from the slope measured on a larger separation range.

\begin{figure}
\centering
\includegraphics[angle=0,width=0.95\hsize]{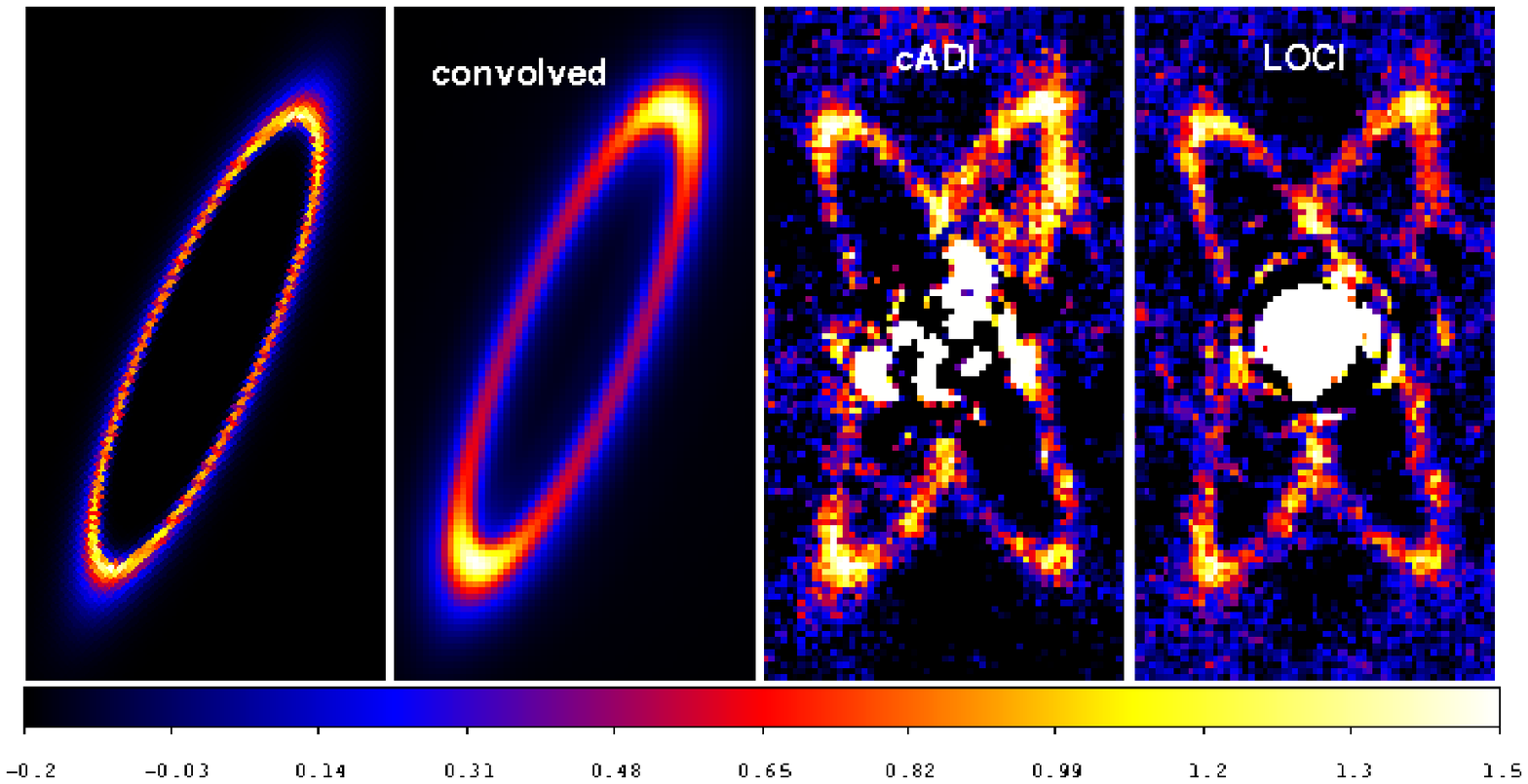}\\
\includegraphics[angle=0,width=0.95\hsize]{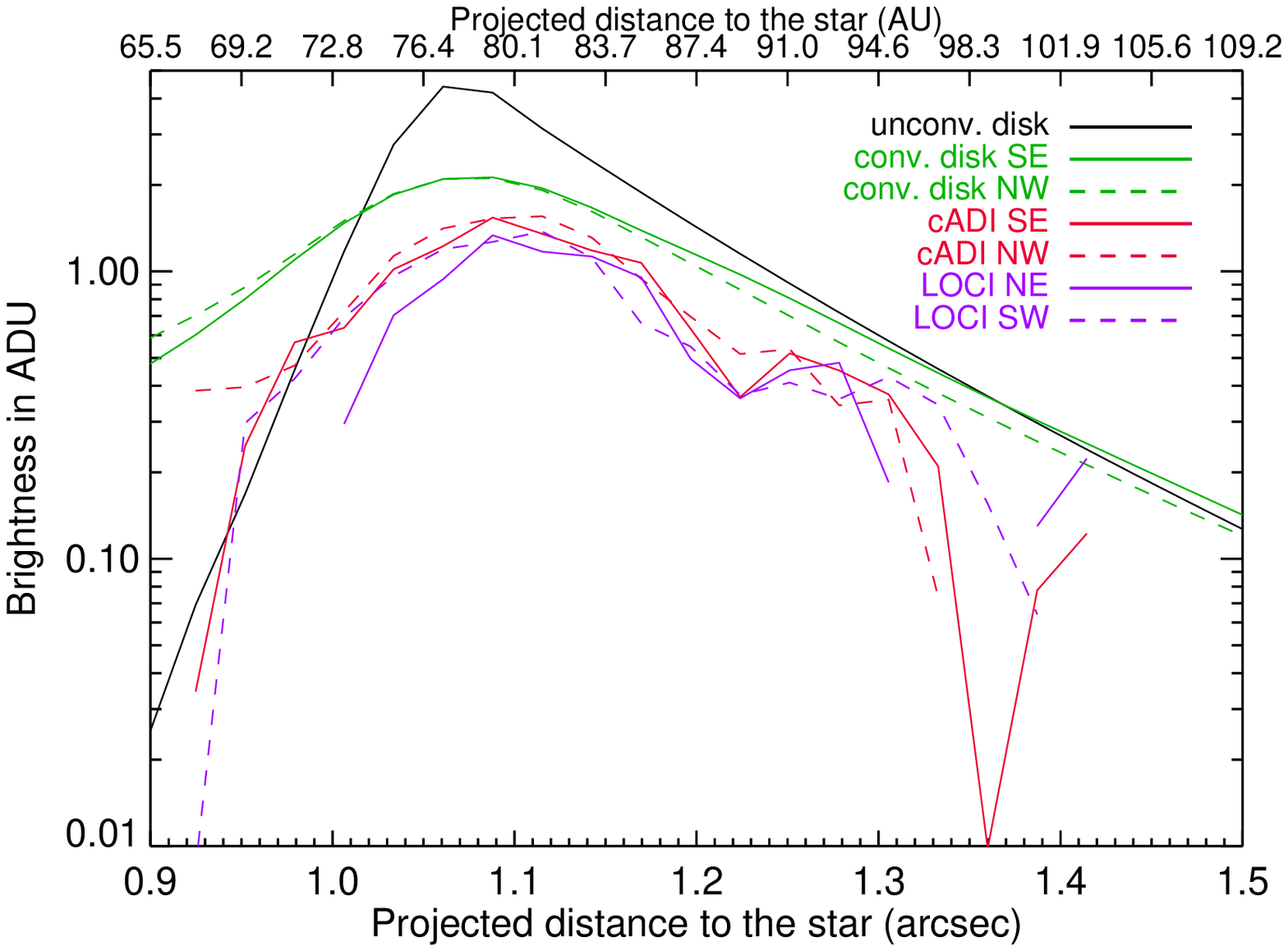}\\
\caption{Top: from left to right: simulated disk, simulated disk after convolution,  after cADI, and after LOCI reductions. Bottom: corresponding SBDs, showing the evolution of the profile after different steps of reduction. See text.}
\label{sbdsteps}
\end{figure}

\begin{figure*}
\centering
\includegraphics[angle=0,width=0.45\hsize]{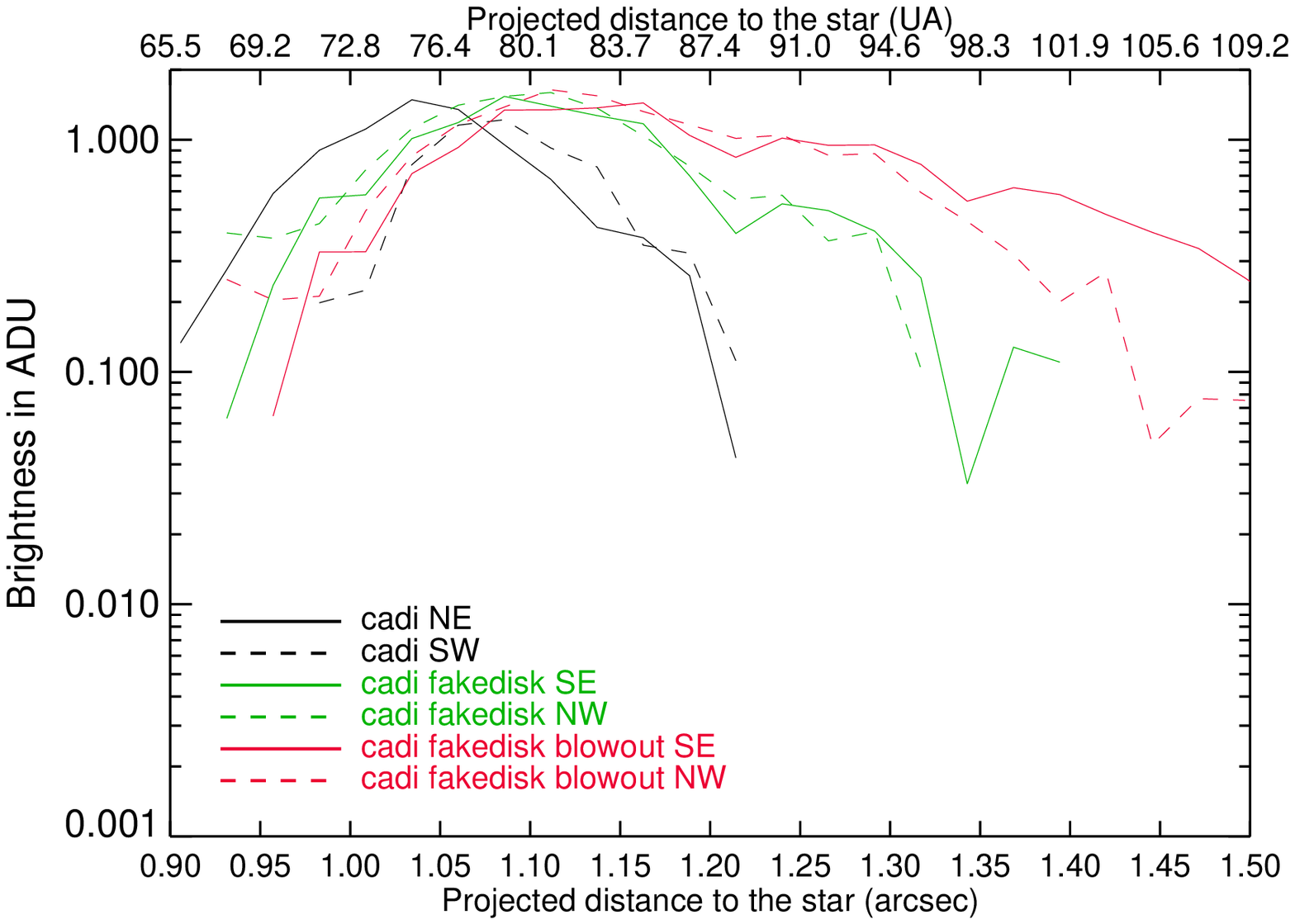}
\includegraphics[angle=0,width=0.45\hsize]{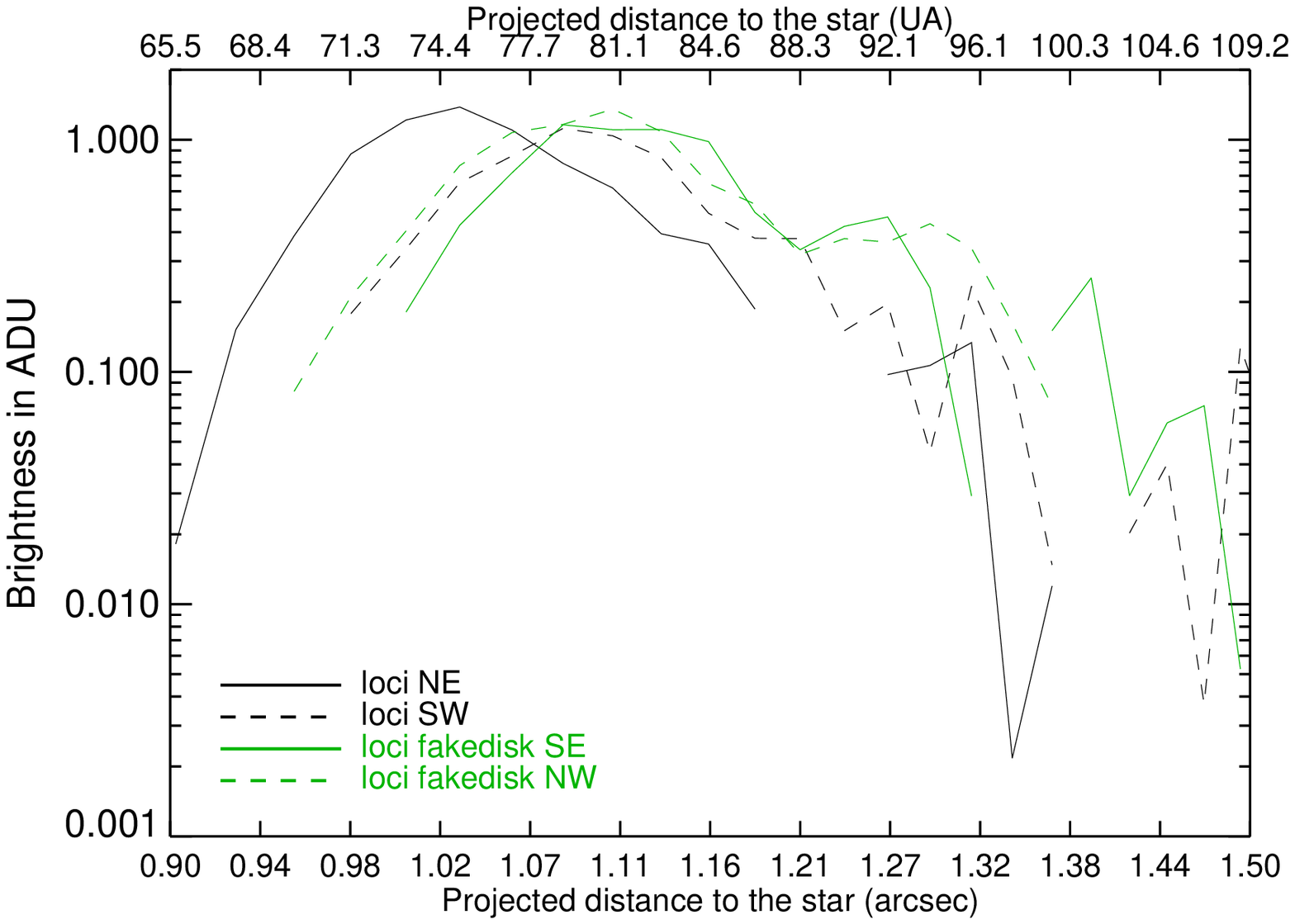}
%obs_profiles_loci_fakediskinjected_blowout_nofit.eps}
 \caption{Left: simulated radial brightness distributions for the HR4796ASD disk (green) and HR4796ASBD disk with blowout (red) at L' along the major axis (log scale) once inserted in the data cube and after cADI reduction has been applied. For comparison, observed SBD (black). Right: simulated radial brightness distributions for the HR4796ASD disk (green) at L' along the major axis (log scale) once inserted in the data cube and after   LOCI reduction has been applied. For comparison, observed SBD (black). }
\label{profiles_disks_noise}
\end{figure*}

The ring shape seems nevertheless to indicate a sharp outer edge, but we need to check the impact of the ADI reduction procedure on the final shape of the disk. Figure~\ref{sbdsteps} illustrates the evolution of the SBD along the semi-major axis, starting from the fake disk HR4796SD, then once the disk is inserted in the real data cubes at 130\deg  (see the corresponding images in Figure~\ref{planche_disksimu_noise}), convolved by the observed PSF, and finally when the datacube is reduced with cADI and LOCI. The SBD shape is clearly impacted. We note that the effect is stronger in the inner region that in the outer one. To test whether we can discriminate between a steep and a less steep outer profile, we  consider the fake disks HR4796SD and HR4796blowoutSD inserted in the real data cubes at 130\deg  (see the corresponding images in Figure~\ref{planche_disksimu_noise}), and convolved with the real PSF. The SBD profiles after convolution and reduction along the semi-major axis are given in Figure~\ref{profiles_disks_noise}. We note that the slight shift between the observed and simulated disks SBD is due to an unperfect assumption on the ring position, and is not relevent here. The observed SBD profile appears to be more similar in shape to the ones corresponding to the HR4796SD case rather than the HR4796blowoutSD one. We conclude then that even when taking into account the possible biases, the data indicate a very steep outer edge, compatible with  $\alpha$$_{out}$ = -10, as found by \cite{thalmann11} rather than a less steep one. We cannot provide precise values to the outer NE and SW slopes with the present data, but they are in any case  different from the  ones measured in the case of the other A-type stars such as $\beta$ Pictoris (typ. between -4 and -5; \cite{goli06}) and HD32997 (\cite{bocca11}), and that 
are expected from a disk which outer part is dominated by grains blown out by radiation pressure from an A-type star (see below). 

\subsubsection{Disk width}
 Our data indicate a ring width of about 0.154" after cADI  reduction for the NE side and 0.136" for the SW side (data binned over 5 pixels). With mcADI, these values are only marginally changed: 0.147" and 0.135" respectively. However, as seen above, the SBD, especially inner to the ring is impacted by the PSF convolution, the amplitude of the FoV rotation, the ADI procedure, the binning used for the extraction of the SBD, the noise level as well as the zero flux level after reduction. We made several test with fake disks to estimate the impact of these steps on the FWHM. Also, we tested the impact of the evaluation of the zero level after reduction. It appears that the disk width is mainly affected in the present case by the PSF convolution and the zero level. Taking all these parameters into account, we cannot conclude that the disk is significantly narrower than the size found by Schneider et al. (2009) 0.197", which once corrected from the broadening by the STIS PSF became 0.18" (13 AU) at  shorter wavelengths.

\section{Companions around HR4796A}
\subsection{SAM detection limits}
The detection limits from the SAM dataset are derived from a 3D
$\chi2$ map. This map has on each axis the three parameters used to
model a binary system: the separation, the position angle, and the
relative flux. This model, fitted on the closure phase, is detailed
in \cite{lacour11}. Visibilities are
discarded. Fig.~\ref{limdet2D_hr4796A} is showing the detection limits
as a function of right ascension and declination. It is obtained by
plotting the 5\,$\sigma$ isocontour of the 3D map (the isocontour
level is given by a reduced $\chi^2$ of 25).

We did not account for the presence of the disk in the model
fitted. We considered that it did not affect the visibilities (because
very faint), and did not affect the closure phase (because
quasi point-symmetric). Nevertheless, it is not impossible
that some of the structures present in Fig.~\ref{limdet2D_hr4796A} may
be caused by a second order effect of the disk on the closure
phase. Thus, neglecting the influence of the disk means that we are
conservative on the detection limit map.

 Given these values,  and assuming V-L' = 0 for this A0-type star, and an age of 8 Myr for the system, we derive the 2D detection limits expressed in Jupiter masses, using DUSTY models (Fig.~\ref{limdet2D_hr4796A}; right). At a separation of about 80 mas
(6 AU), we exclude the presence of companions with masses larger than 29 \mjup. At a
separation of 150 mas, the limit becomes M = 40 \mjup (DUSTY). In both cases, COND03 models give similar limits within 1 MJup.  At 60 mas, the detection limit is M = 50 \mjup (DUSTY) and would be 44 MJup with COND03. Such values fall in the mass range of brown dwarfs and represent unprecedented mass limits for this range of separations.

\begin{figure*}
\centering
\includegraphics[angle=0,width=.45\hsize]{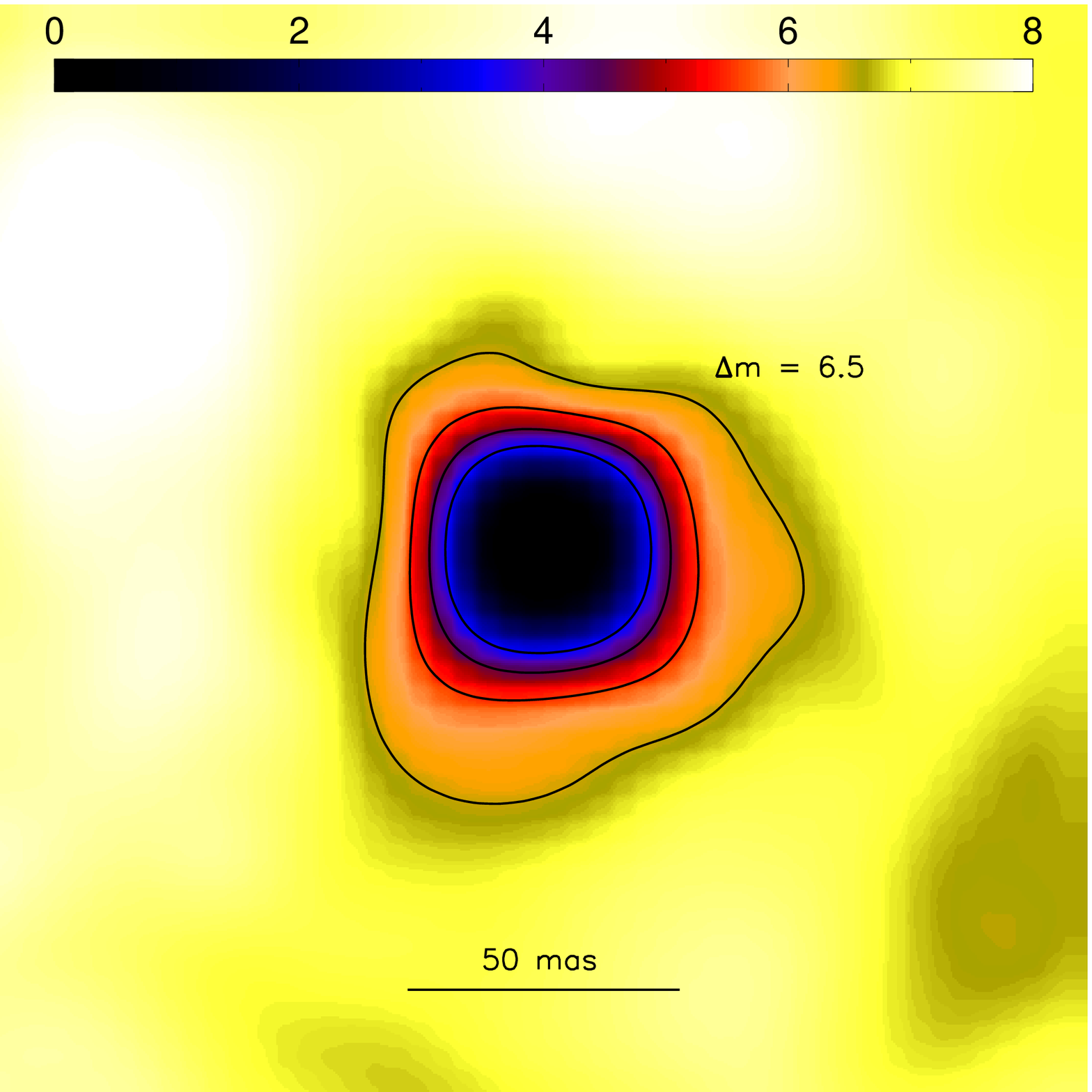}
\includegraphics[angle=0,width=.45\hsize]{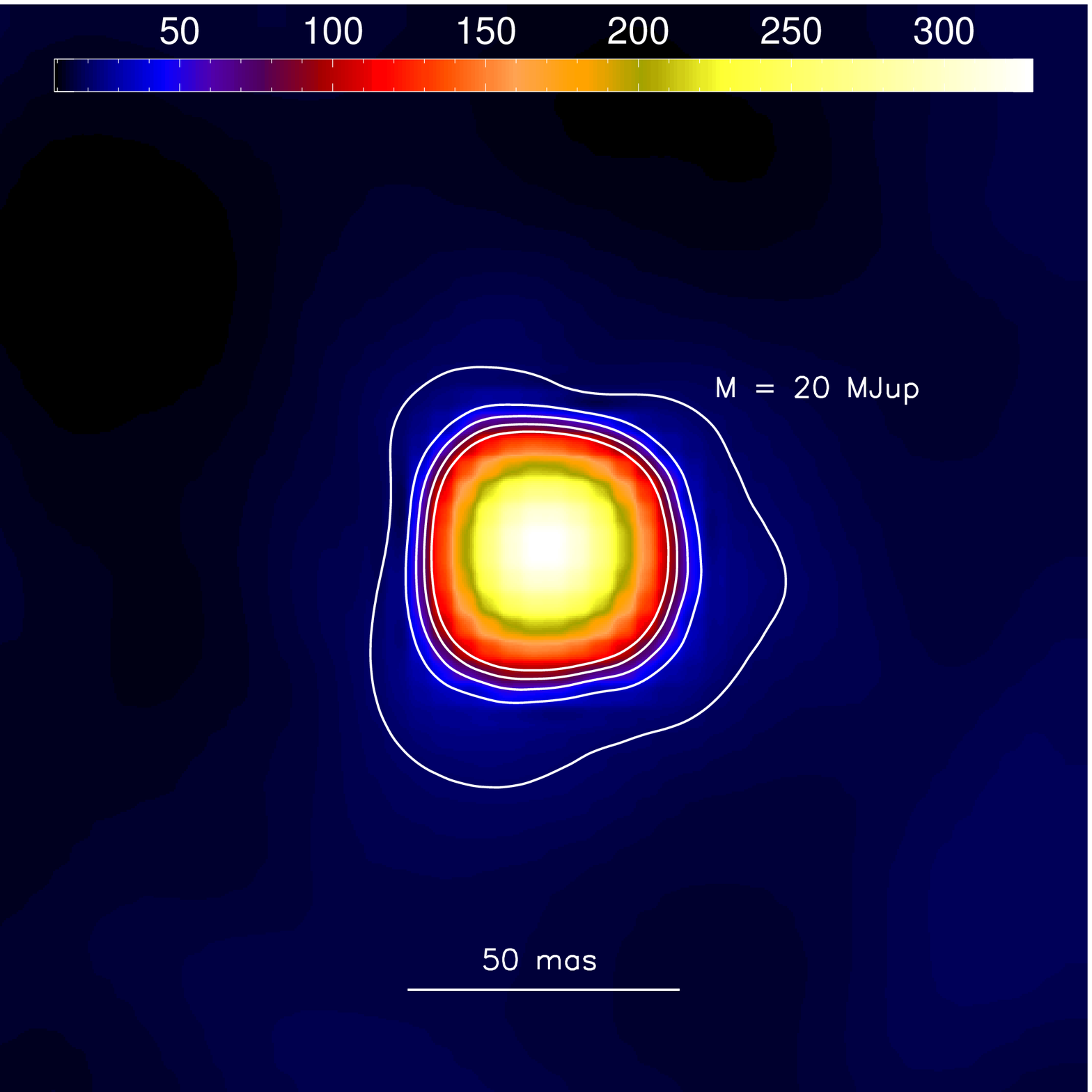}
\caption{Left: 2D map of 5-$\sigma$ detection limit of point-like structures towards HR4796A, using SAM data. Isocontours for $\Delta$mag = 3.5, 4.5, 5.5, and 6.5. Right: detection limits expressed in Jupiter masses using COND03  models; isocontours for masses of 20, 40, 60, 80 and 100 MJup.  In both images, North is up and East to the left.} 
\label{samlimdet}
\end{figure*}

\begin{figure}
\centering
\includegraphics[angle=0,width=.95\hsize]{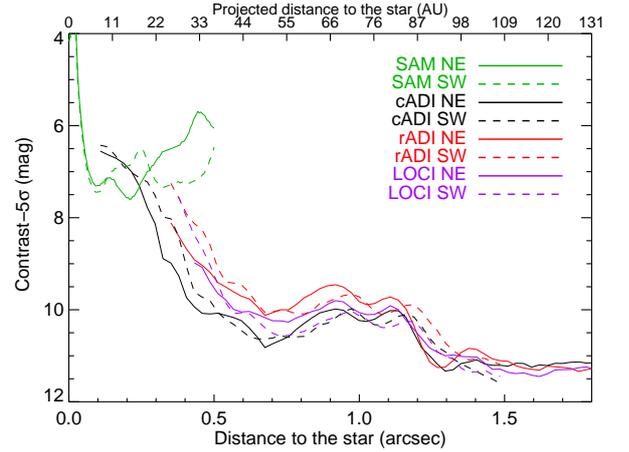}
%detlim_cadi_sadi_radi_loci_PA26.ps}
  \caption{5-$\sigma$ detection limit of point-like structures around HR4796A, along the major axis in the NE and SW directions using SAM and ADI data.}
\label{limdet1D_hr4796A_PA26}
\end{figure}

\subsection{ADI detection limits}
\begin{figure*}
\centering
\includegraphics[angle=0,width=.45\hsize]{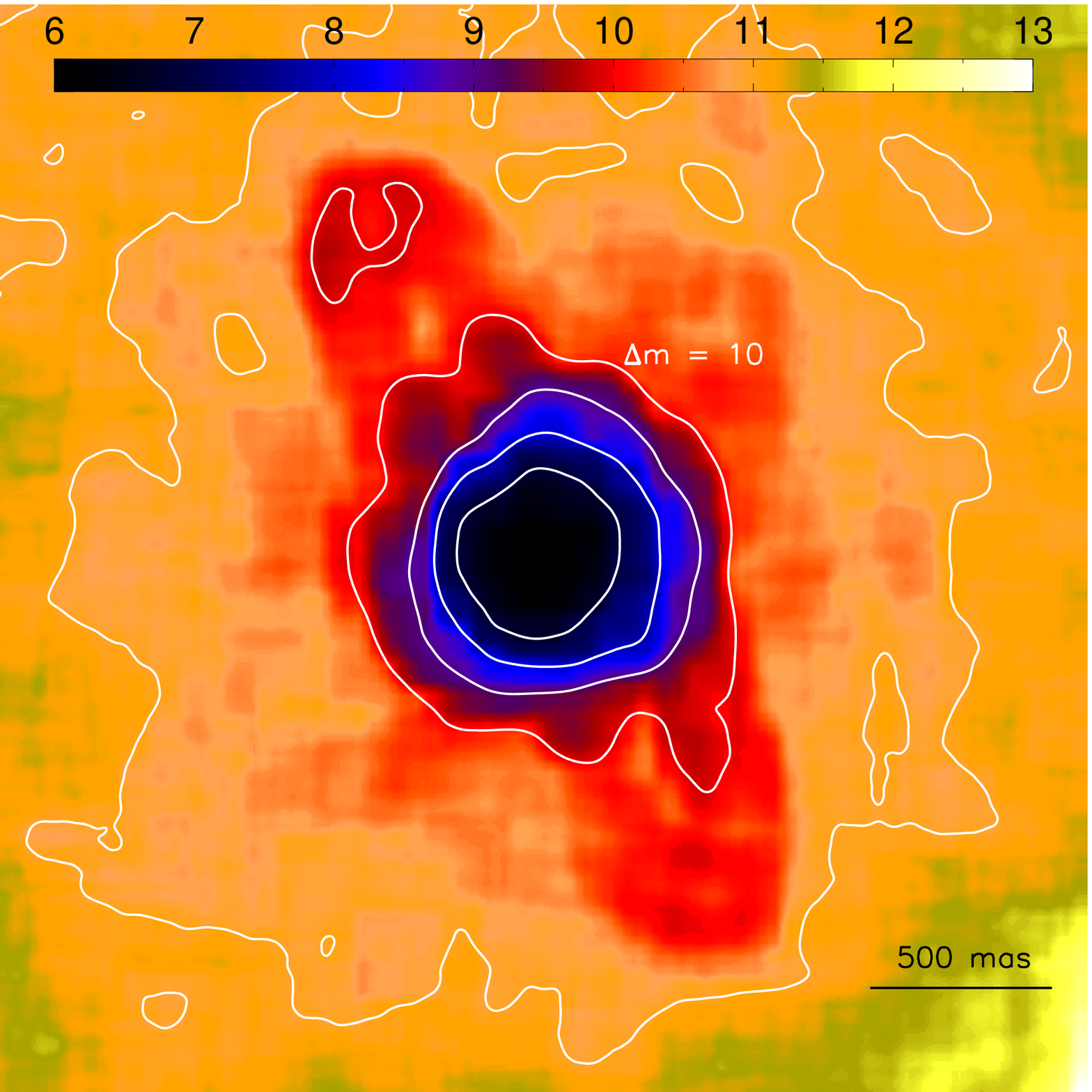}
\includegraphics[angle=0,width=.45\hsize]{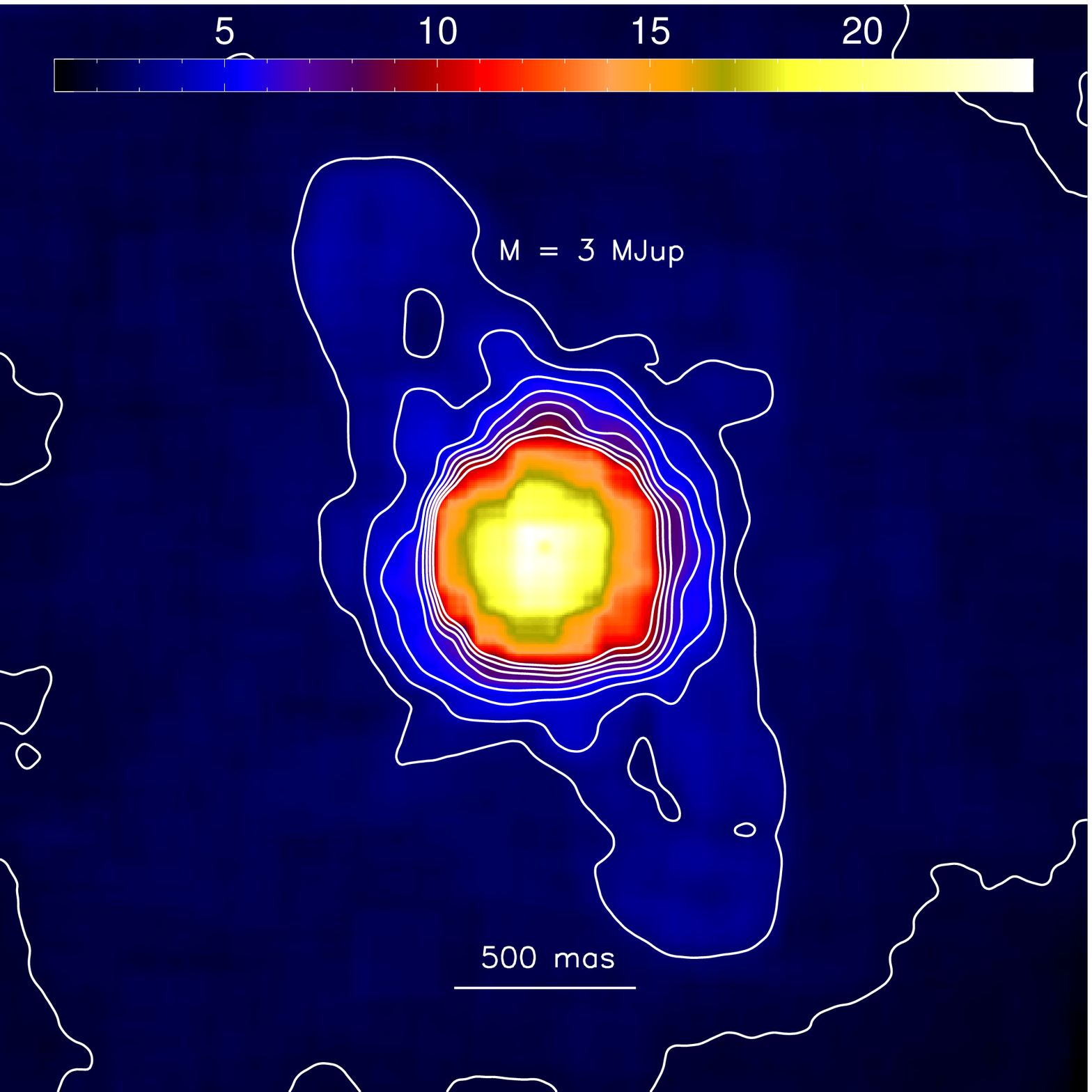}
  \caption{2D map of 5-sigma detection limit of point-like structures towards HR4796A, using all available data (CADI). Left: expressed in contrasts (isocontours from 7 to 11 mag., steps of 1 unit). Right: translated in masses using COND03 models (isocontours from 2 to 10 \mjup, steps of 1 \mjup).}
\label{limdet2D_hr4796A}
\end{figure*}

\begin{figure}
\centering
\includegraphics[angle=0,width=\hsize]{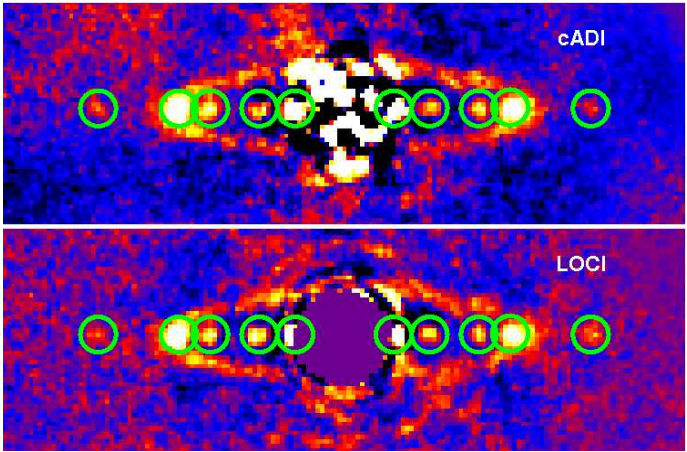}
  \caption{Fake planets inserted at 0.3, 0.5, 0.8, and 1" on both sides of the disk, with fluxes corresponding to the 5-$\sigma$ detection levels, and reduced with cADI and LOCI.}
\label{FP}
\end{figure}

We computed the detection limits using the data obtained on April 6th and 7th, with different reduction methods. To estimate them, we took into account the flux losses due to the ADI reductions, either injecting fake planets in cubes of empty frames obtained with similar FoV rotation for the cADI procedure, or injecting fake planets in the real data cubes in the case of the LOCI procedure. The noise was estimated using a sliding 9 pixel wide box along any given PA. We checked the obtained detection limits by inserting fake planets with fluxes corresponding to the 5$\sigma$ limit at different separations, and processed again the data, and mesured the resulting S/N ratio on the planets. The S/N ratio  were close (or sometimes slightly larger than 5) which shows that our limits are properly estimated. The 2D-detection limits are shown in Figure~\ref{limdet2D_hr4796A} for cADI, expressed either in contrasts or in masses, using the COND03 models (\cite{baraffe03}) or BT-settl models (\cite{allard11}) and assuming an age of 8 Myr. Similar (not better) limits were obtained with rADI and LOCI. To check the robustness of the detection limits obtained, we injected fake planets with fluxes corresponding to the 5$\sigma$ level as well as the fake disks and processed the data cubes as before. The resulting images (see Fig.~\ref{FP}) revealed the planets  with at least a 5$\sigma$ level. 

The 1-D limits along the major axis, at a PA of 26 degrees (NE side of the disk) are showed in Figure~\ref{limdet1D_hr4796A_PA26}. Similar values are obtained in the SW direction. A few values expressed in jovian masses are given in Table~\ref{limdet}.
 The detection limits are better than the SAM ones further than $\simeq$ 0.25-0.3", with a value of  about 7.5 \mjup at 0.25-0.3";  they are below 3.5 \mjup for separations in the range 0.5-1", and well below further than 1.5". Alltogether, these  are to our knowledge the best detection limits obtained in the close surroundings of HR4796A.

\begin{table}[t!]
\caption{ A few detection limits along the semi-major axis of HR4796A (NE side). First lines (top) values are derived from SAM data, using DUSTY models (note that COND models agree within 1-2 MJup in most cases). The other values (bottom) are derived from ADI data, using COND03 or BT-SETTLE models.}
\label{stats}
\begin{center}
\begin{tabular}{ l l l l l l l l l}
\hline 
 Sep (")& Sep (AU)& $\Delta$mag (5$\sigma$) &Mass (\mjup) \\ 
\hline
0.06"& 4.5& 6.9 & 50 (DUSTY)\\
0.08"& 6& 7.3& 29 (DUSTY/COND)\\
0.15"&11&7.0&40 (DUSTY/COND)\\
0.3" & 22& 7.2& 32 (DUSTY/COND)\\
\hline
0.3" & 22& 8.2& 7 (BTSETTLE, COND)\\
0.5" & 36.5& 10.1&3.5 (COND)\\
0.8" & 58& 10.34 &  3 (COND)\\
1" &73&10.23 &  3  (COND)\\
%mag_comp,10^(-11.8/2.5),5.8,13.74 => 13.3
1.5" &110& 11.22&   2 (COND)\\
%mag_comp,10^(-12/2.5),5.8,13.74 => 13.5
     \hline
    \end{tabular}
  \end{center}
\label{limdet}
\end{table} 

\section{Companions around HR4796B}
\begin{figure*}
\centering
\includegraphics[angle=0,width=.45\hsize]{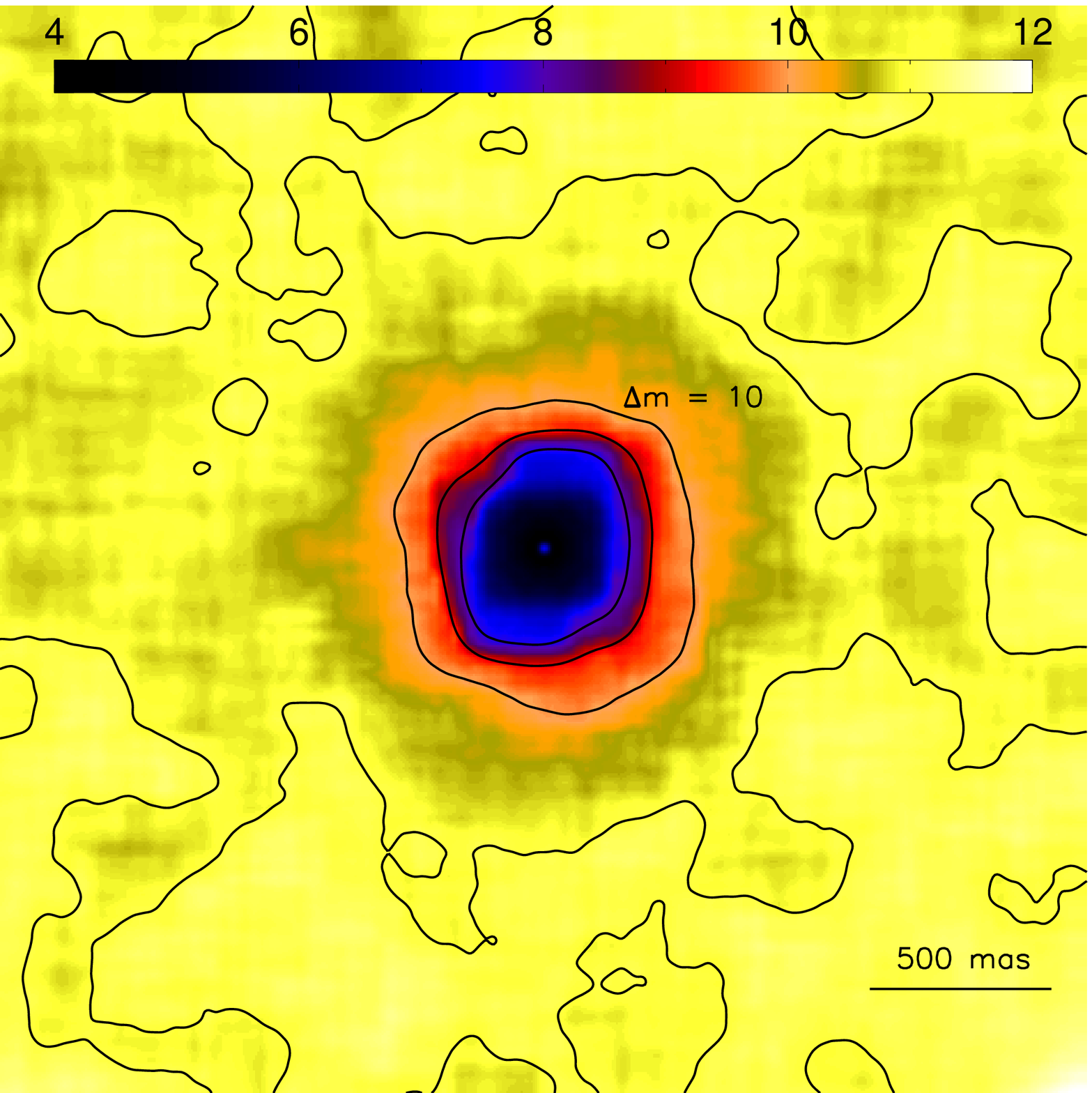}
\includegraphics[angle=0,width=.45\hsize]{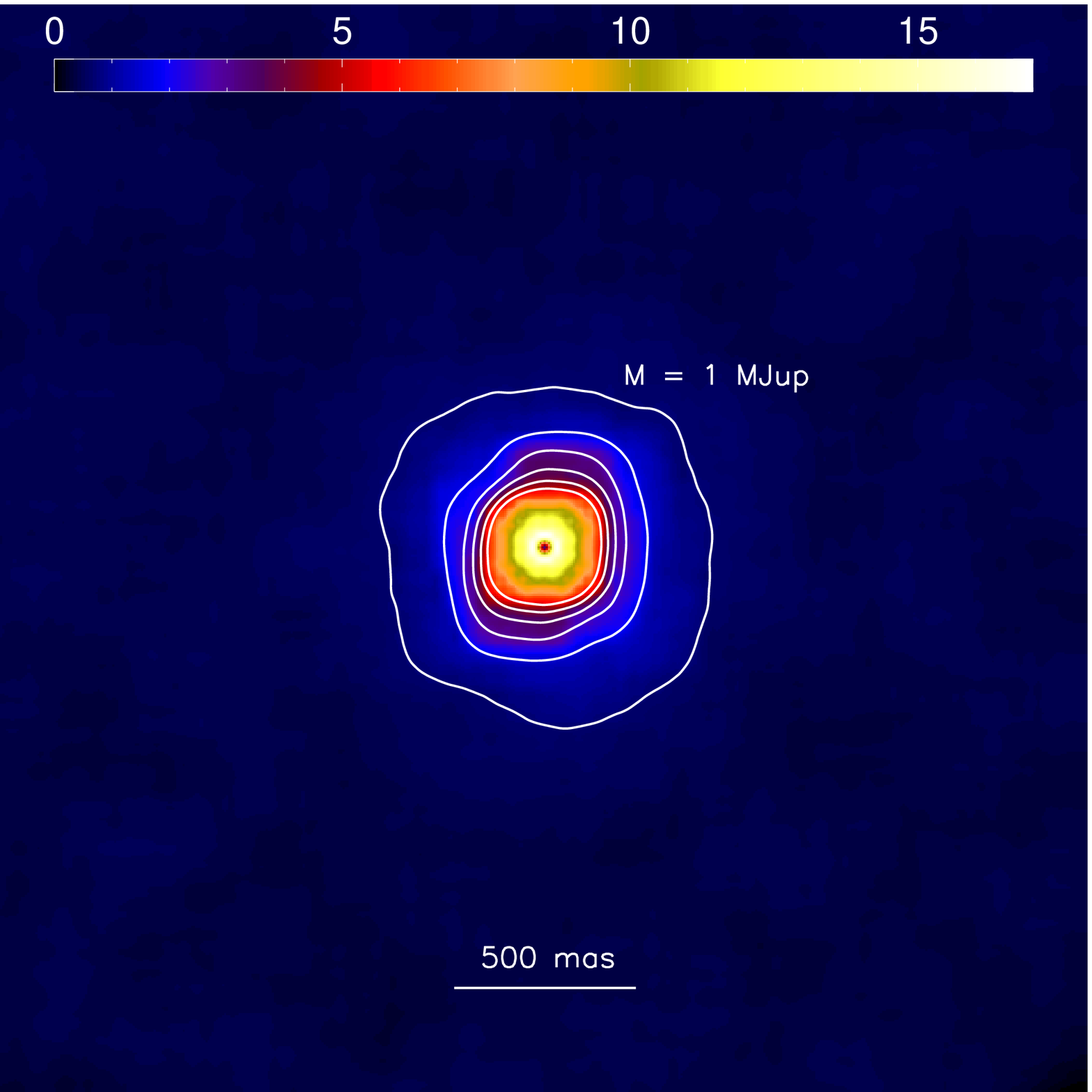}
%detlim_map_HR4796B.jpg
  \caption{2D map of 5-sigma detection limit of point-like structures around HR4796B (April 6th), expressed in contrast (Left), with contour-levels of 8 to 11 mag (step =1 mag); and in Mass (Right), with contour-levels of 1 to 6 MJup (step =1 MJup).}
\label{limdet2D_hr4796B}
\end{figure*}

HR4796B has the following magnitudes: V = 13.3, H = 8.5, K = 8.3. Using the observed contrast between HR4796A and HR4796B  (2.6 mag) on the present data, we find L'$\simeq$ 8.4 for HR4796B, hence L'abs $\simeq$ 4.1. This value is in agreement with the Lyon's group model (\cite{baraffe03}), which, given the near-IR colors, predicts an absolute L' magnitude of 3.8. We give in Figure\ref{limdet2D_hr4796B} the 2D map of the detection limits (5$\sigma$), both in terms of contrast magnitudes and Jupiter masses. At 0.3", masses as low as 2 \mjup could be detected, and at 0.5", the detection limit is below 1\mjup. 

\section{Discussion}
\subsection{The inner disk sharp edge}

One of the most remarkable features of the HR4796 disk is certainly the offset of the disk center with respect to the star (also observed in the case of HD141569, and Fomalhaut). Two explanations are 1) the presence of a close, fainter companion (in such a case, the disk would be a circumbinary disk and orbit around the binary center of mass), and 2) the presence of a companion close to the disk inner edge on an eccentric orbit that induces a forced eccentricity to the disk ring by secular gravitational interaction, an explanation which was proposed to explain the eccentricity of the Fomalhaut disk (\cite{quillen06}, \cite{kalas05}, \cite{kalas08}, \cite{chiang09}). 

We first investigate whether this offset could be due to the presence of a close companion. In such a case, the ellipse center would mark the center of mass of the binary system. Using the center of mass definition, it appears that the mass of a body necessary to shift the center of mass at the observed position of the ellipse center would be much larger than the detection limit obtained with SAM between 40 and 400 mas or between 400 mas and 1 arcsec (ring position) with the ADI data. A companion located between 23 and 40 mas would have a mass larger or comparable to that of HR4696A; such a scenario must be excluded as under such conditions, the photometric center of the system would also be shifted. Then, the most plausible explanation to the offset is a light eccentric planet close to the inner edge of the disk. 

\begin{figure}
\centering
\includegraphics[angle=0,width=.9\hsize]{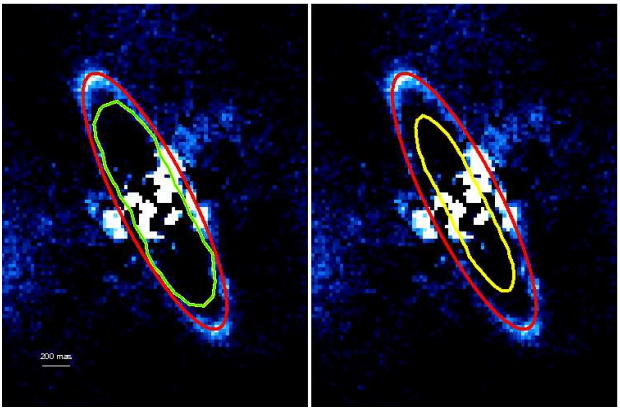}
  \caption{ Possible remaining locations of a planet inner to the disk that could produce an inner sharp edge at 77 AU given the present detections limits (Left: case of a circular orbit; Right: case of an elliptical orbit with e=0.1). The possible locations are restricted to the region between the yellow curve and the red ellipse that mimics the disk itself. North is up, East to the left. }
\label{map_planete}
\end{figure}

We now try to use the detection limits found in this paper to constrain the properties of an inner planet that could be responsible for the steep inner edge observed with HST/STIS data, which, conversely to ADI data, is not impacted by ADI reduction effects. \cite{wisdom80} showed that in the case of a planet and particles on  circular orbits, we have the relation $\delta$a/a = 1.3.(Mp/Ms)$^{2/7}$ where Mp and Ms are the planet and star masses, a is the orbital radius of the planet and $\delta$ a the distance between the planet and the disk inner edge. Hence if a planet sculpts the inner edge, its mass and distance from the inner edge must satisfy this relation. Assuming an inner edge located at 77 AU, we can derive the mass of the planet necessary to produce this sharp edge, as a function of its distance to the edge, and test whether such a planet would have been detected or not. This is done in Figure\ref{map_planete} where we show the region, inside the yellow ellipse, that, given the present detection limits, have to be excluded. Hence the only possible location of the planets responsible for the inner edge is between the yellow ellipse and the red one (which traces the inner edge of the disk).
% ; we also show for illustrations the isocontours for M=1.5 (outer curve), 2.5 and 7.5 (inner curve) \mjup. Such masses correspond to contrasts of about 11.5-12, 10.5 and 8.3. Comparing Figure\ref{map_planete} and Figure\ref{limdet2D_hr4796A}, 
We see that  along the major axis, only the planets closest to the inner edge (less than $\simeq$ 10 AU) remain out of the present detection capabilities. Hence if a planet is responsible for the inner edge sculpting and is located along or close to the major axis, then it should be a low mass planet, and located further than 63 AU, ie within about 15 AU from the edge. Along the minor axis, due to the projection effects, the presence of planets is much less constrained: only planets at more than 26 AU from the edge would have been detected.  

The previous constrains were obtained assuming the planet and the perturbated bodies are both on circular orbits. The actual disk eccentricity beeing very small, about 0.02, this assumption is reasonable. As an exercice, we investigate the impact of a higher eccentricity, using the results of \cite{mustill11}, who revisited this scenario, assuming the perturbating bodies were on an eccentric orbit; the relation becomes : $\delta$ a/a = 1.8.e$^{1/5}$.(Mp/Ms)$^{1/5}$. With the same reasonning, and assuming an eccentricity of 0.1, we provide in Figure\ref{map_planete} the possible locations of a planet responsible for the inner edge, with the same color conventions as in the circular case. Again, comparison with Figure\ref{limdet2D_hr4796A} shows that if a planet was responsible for the inner edge sculpting, and located along or close to the semi-major axis, then it would have to be located less than $\simeq$ 25 AU from the edge of the disk. The location of planets along or close to the minor axis would not be significantly constrained. This case, even though not adapted to the present case as the disk eccentricity is very small, illustrates the impact of this parameter on the planet  detection capabilities.

\subsection{The outer disk sharp edge}
 Another striking feature of the system is the very steep disk outer edge.  Radiation pressure from A-type stars induces surface brightness distributions in the outer part of disks with slopes of -3.5 to -5, depending on the assumptions related to the production laws of small grains (see for instance \cite{leca96}, \cite{thebault08} and ref. there-in).  In particular, \cite{thebault08} modeled  the outer parts of collision rings with an initially steep outer edge, following the motion of the small grains produced through collisions and submitted to radiation pressure and showed that, as already proposed, the profile of the resulting SBD in the outer part of the disk followed a r$^{-3.5}$ law. They showed that unless the disks are extremely and unrealistically dense, and prevent the small grains from escaping, the disks have to be extremely "cold" (with an average free  eccentricity of $\leq$ 0.0035) to explain an outer power law of r$^{-6}$ (which was the value adopted at this time for the HR4796 profile). AO data suggest that the situation could be even more radical with an even steeper outer edge than previously thought. Also, if confirmed, the fact that we possibly find at 3.8 $\mu$m a disk width different from that found in the optical (0.2-1$\mu$m) with STIS would argue against an extremely dense disk, as, in such a case, large bodies and small grains would be in the same regions. The cold disk scenario can nonetheless be also problematic as, in such a case small grains should be underabundant (\cite{thebault08}) and the optical/near-IR fluxes would be produced  by  large particules (typ. sizes 50 $\mu$m). We should expect then a color index different from that observed (see \cite{debes08}). The latter rather predict that the scattered light flux is dominated by 1.4 $\mu$m dust, which seems difficult to explain within the dynamically-cold-disk scenario. 

Could gas be responsible for the observed steep outer edge? 
  \cite{takeuchi01} investigated the impact of the gas in such debris disks, and showed that even small amount of gas, 1- a few Earth masses, could partly balance the effects of gravitation, radiation pressure and Poynting-Robertson drag and alter the grains dynamics differentially, and lead to grains spatial distributions, that, depending on the grain sizes, could be different from those expected in a disk-free gas. Under such processes, the gas could be responsible for ring-like structures at distances depending on the dust size considered. In their attempt to investigate disks roughly similar to HR4796 and HD141569, assuming 1 MEarth gas, they showed that, conversely to large grains which occupy the whole gas disk, grains with sizes ($\simeq$ 10-200 $\mu$m) tend to concentrate in the outer gas disk, where the gaseous density sharply decreases. Hence these grains would form a narrow ring which position traces the change in the radial distribution the gaseous disk. Under this {\it a priori} attractive scenario, grains with sizes 1-10 $\mu$m would still be blown away.
 The main problem with this hypothesis is that it requires a gas disk with a relatively sharp outer edge, and thus an explanation for such an edge. Another issue is that \cite{takeuchi01} did not explore the SBD profile beyond the main dust ring, so that it remains to be see whether  slopes in the -10 range are possible. 
Finally, it is worth noting that so far no circumstellar gas has been detected  either in atomic species through absorption spectroscopy (but the system being inclined, the non detection is not a strong constraint) or molecular species, either CO (\cite{liseau99}) or H2 (N(H2)$\leq 10^{15} cm^{-2}$ \cite{martin08}). In any case, such a scenario can be tested in the forthcoming years with high angular resolution observations on a wide range of wavelengths.

We finally study the possibility that the outer disk is sculpted by massive bodies. The first candidate we might think of is HR4796B. \cite{thebault10} investigated the possibility that the disk could be sculpted by HR4796B, if orbiting on a rather eccentric orbit (e$\geq$ 0.45) but again showed that, even under such conditions the outer profile would not be so steep. This is mainly because the companion star is not able to dynamically remove small grains from the outer regions at a pace that can compensate for their steady collisional production in the parent body ring.

An alternative explanation could be the presence of a close, unseen outer planet. 
We investigate this scenario using the new code developed by \cite{thebault2012} to study perturbed collisionally active debris discs. The code computes the motion of planetesimals submitted to the gravitational perturbation of a planet; and follows the evolution of small dust realeased through collisions among the planetesimals and submitted to radiation pressure and Poynting-Robertson effect (note that in the present case, radiation pressure largely drives the grains dynamics once produced). The configuration we consider is a narrow ring of large parent bodies, a birth ring of width $\sim 8$AU centered on 71 AU. The collisional production and destruction rates of small grains (those which contribute to the luminosity beyond the main ring) in this parent body ring is parameterized by the average vertical optical depth within the main ring, taken to be $\tau = 5\times 10^{-3}$  \footnote{for more detail on the procedure, see \cite{thebault2012}}. The resulting SBD (case face-on) is derived  assuming grey scattering. It can be directly compared to the curve presented in Figure 6 of Schneider et al (2009) paper (de-projected curve\footnote{the FWHM of the STIS PSF being quite narrow compared to the disk width, its impact is very limited.}). 

For the perturbing planet's mass, we consider 3 different values:  $8M_{Jup}$, $5M_{Jup}$ and $3M_{Jup}$, which are consistent with the constraints imposed by our observational non-detection. Note that $8M_{Jup}$ is only marginally possible in a very narrow region along the disc's semi-minor axis, but we have to keep in mind that the detection limits are derived from masses-brightness relationships that are debated at young ages. 
We assume a circular orbit for the planet (the most favourable case for cleaning out the region beyond the ring, see \cite{thebault2012}) and place it as close as possible to the parent body ring, i.e., so that the outer edge of the observed ring (around 75 AU) corresponds to the outer limit for orbital stability imposed by planetary perturbations.  This places the planet at a distance to the central star comprised between $92$ and $\sim 99$ AU depending on its mass.

\begin{figure*}
\centering
\includegraphics[angle=0,width=.45\hsize]{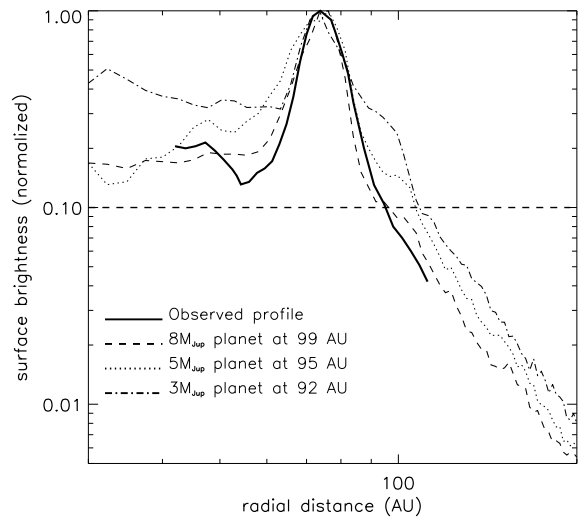}
\caption{ Synthetic surface brightness profiles obtained, using Thebault (2012)'s numerical model, for 3 different masses of a putative perturbing outer planet: $8M_{Jup}$, $5M_{Jup}$ and $3M_{Jup}$. For each case, the planet has a circular orbit and is placed as close as possible to the main ring of large parent bodies in order to truncate it at about 75AU without destroying it (see text for more details).
The observed profile, derived from \cite{schneider09}, is shown for comparison. For each planet mass, we show the radial cut (i.e., for one position angle along the disc) that provides the best fit to this observed profile.
The horizontal line delineates approximately the part (above this line) of the SBD accessible to the observations assuming a dynamical range of 10.
}
\label{dyn_simu}
\end{figure*}

In Figure~\ref{dyn_simu}, we show the SBD obtained for such a configuration. Note that, for each planet mass, we are not showing an azimutal average but the "best" radial cut, i.e. the one that gives the closest match to the deprojected NE side SBD obtained in Figure 6 of \cite{schneider09}.
As can be clearly seen, the $8M_{Jup}$ case provides a good fit to the observed profile: the maximum of the SBD roughly corresponds to the outer edge of the parent body disk and is followed by a very sharp brightness decrease, with a slope $\leq$ -10 between 75 and 95 AU, i.e. between brightnesses of 1 and 0.1, a range corresponding approximately to the dynamical range accessible to the available images.  This is significantly steeper than the one that would be  expected if no planet was present (-3.5 according to \cite{thebault08}) and is fully compatible with the observed sharp luminosity decrease.

Longwards 95 AU, the flux level is lower ($\leq$ 0.1 ADU) and the SBD is flatter (slope $\simeq$ -3.8). 
%The slope of the SBD measured using all the separation range available longwards the maximum is found to be -6.9.
 We also note a plateau inside the parent body ring at a level of $\sim 0.2$, due to both the inward drift of small grains because of the Poynting-Robertson effect and to the dynamical injection of particles after close encounters with the planet. Of course, not too much significance should be given to the SBD obtained inwards of the disc since our simulations (focused on the outer regions) do not consider any inner planet shaping the inner edge of the disc. Nevertheless, they show that, should "something" have truncated the disc at around 67\,AU in the past, then the effect of one external planet on such a truncated disc could lead to an SBD compatible with observations in the inner regions.
For the $5M_{Jup}$ case, the fit of the observed SBD is slightly degraded, but mostly in the region beyond 90\,AU where flux levels are close to the 0.1 threshold. For the $3M_{Jup}$ case, however, the fit gets very poor for almost the whole outer region (keeping in mind that we are here showing the best radial cut).

We conclude that a $8M_{Jup}$ planet located on a circular orbit at $\sim 25\,$AU from the main ring provides a satisfying fit (especially considering the non-negligeable uncertainties regarding flux values far from the main ring) to the observed SBD. According to the derived detection limits, such a massive planet would have been detected almost everywhere except in a very narrow region along the disc's semi-minor axis; however, 
we remind the uncertainties inherent to the models used to link planet masses and luminosities as a function of the system's age.
In any case, even a less massive perturber of e.g. $5_{MJup}$ would still give an acceptable fit of the observed luminosity profile. The external planet scenario thus seems the most likely one for shaping the outer regions of the disc.
Of course, these results are still preliminary and should be taken with caution. A more thorough numerical investigation should be carried out, exploring a much wider parameter space for planet masses and orbit, as well as deriving other outputs that can be compared to observations, such as 2-D synthetic images. Such a large scale numerical study exceeds the scope of the present work and will be the purpose of a forthcoming paper.
Note also that the more general issue of how planets shape collisionally active debris disks will be thoroughly investigated in a forthcoming paper (Thebault, 2012b, in prep.).

\section{Summary and future prospects}
In this paper, we have provided the first high-resolution images of the HR4796A disk at L' band. They allow us to see a narrow disk at almost all PA. As the technics used, Angular Differential Imaging is expected to impact the final disk shape and appeareance, we have developped simulations to investigate quantitatively the impact of the reduction procedures on the disk parameters. We conclude that the information on the inner part of the disk is significantly impacted, and that the procedure may in some cases (depending on the amplitude of rotation of the field of view), produce important artefacts. This is specially true for LOCI reduction, while classical ADI affects the data to a lower extent. We showed in particular that the streamers detected by \cite{thalmann11} at the outer edge of the disk are probably due to such artefacts. 

Using both ADI and SAM data, we have derived unprecedented lower limits to the presence of planets/companions down to 25 mas from the star. The present data allowed then to put first interesting constrains on the location of the possible planet that could  produce the inner edge of the disk. We showed that the planet responsible for the inner edge must be closer than 15 AU from the ring if located along or close to the semi-major axis. The forthcoming high dynamics instruments such as SPHERE on the VLT and GPI on GEMINI will allow to test this hypothesis with much more accuracy, and be able to actually detect this planet in most cases. 

  We have discussed several hypotheses to explain the sharp outer edge of the disk: gaseous disk, dynamically cold disk, planet on the outer edge. Using detailed simulations, we showed that a planet located  outside the planetesimal ring could nicely reproduce the STIS data. Further simulations will help to better constrain the planet and parent bodies characteristics.

In any case, this work shows how disks characteristics can help constraining possible planet properties. A very important information can be brought by the dependance of the disk properties (ring width, SBD) as a function of wavelength. Resolved images in the future will be crucial to further understand this system.

\begin{acknowledgements}
We acknowledge financial support from the French Programme National de Plan\'etologie (PNP, INSU). We also acknowledge support from the French National
Research Agency (ANR) through project grant ANR10-BLANC0504-01 and ANR-2010 BLAN-0505-01 (EXOZODI). We also thank C. Marois, B. McIntosch and R. Galicher for discussing their L' Keck data with us. We also thank the anonymous referee for his/her helpful comments.
\end{acknowledgements}

\end{document}